%% file: paper.tex
\documentclass[sigconf]{acmart}

\AtBeginDocument{%
	\providecommand\BibTeX{{%
			\normalfont B\kern-0.5em{\scshape i\kern-0.25em b}\kern-0.8em\TeX}}}

\usepackage{tikz}
\usepackage{amsmath}

\usepackage{filecontents}

\usepackage{balance}

\usepackage[inline]{enumitem}
\usepackage{graphicx}
\usepackage{xcolor}
\usepackage{listings}
\usepackage{comment}
\usepackage{subcaption}
\usepackage{url}
\usepackage[htt]{hyphenat}
\usepackage[ruled,vlined,linesnumbered]{algorithm2e}

\usepackage{multirow,tabularx}

\usepackage{makecell}

\usepackage{mdframed}

\setcopyright{acmcopyright}
\copyrightyear{2018}
\acmYear{2018}
\acmDOI{10.1145/1122445.1122456}

\acmConference[Woodstock '18]{Woodstock '18: ACM Symposium on Neural
	Gaze Detection}{June 03--05, 2018}{Woodstock, NY}
\acmBooktitle{Woodstock '18: ACM Symposium on Neural Gaze Detection,
	June 03--05, 2018, Woodstock, NY}
\acmPrice{15.00}
\acmISBN{978-1-4503-XXXX-X/18/06}

\definecolor{pinkshocking}{rgb}{0.99,0.01,0.84}

\definecolor{mGreen}{rgb}{0,0.6,0}
\definecolor{mGray}{rgb}{0.5,0.5,0.5}
\definecolor{mPurple}{rgb}{0.58,0,0.82}
\definecolor{backgroundColour}{rgb}{0.99,0.99,0.99}

\definecolor{mGrayBox}{HTML}{e0e0e0}

\usepackage{ifthen}
\newcommand{\fix}[1]{
	\ifthenelse%
	{\boolean{showcomments}}%
	{\textcolor{red}{#1}}%
	{}%
}

\mdfdefinestyle{HighlightFrame}{
	innermargin     =0.1cm,
	outermargin     =0.1cm,
	linewidth=0.5pt,
	backgroundcolor=mGrayBox,
}

\usepackage[subtle]{savetrees}

\newcommand{\circled[1]}{\tikz[baseline=(char.base)]{\node[font=\sffamily,
shape=circle,draw,inner sep=0.5pt,color=white,fill=black] (char) {#1};}}

\lstdefinestyle{CStyle}{
	backgroundcolor=\color{backgroundColour},
	commentstyle=\color{mGreen},
	keywordstyle=\color{magenta},
	numberstyle=\tiny\color{mGray},
	stringstyle=\color{mPurple},
	basicstyle=\footnotesize,
	breakatwhitespace=false,
	breaklines=true,
	captionpos=b,
	keepspaces=true,
	numbers=left,
	numbersep=5pt,
	showspaces=false,
	showstringspaces=false,
	showtabs=false,
	tabsize=2,
	xleftmargin=2em,
	framexleftmargin=1.5em,
	language=C
}

\usepackage{booktabs}

\usepackage{soul}
\setstcolor{red}

\newcommand{\ie}{\textit{i.e.,} }
\newcommand{\eg}{\textit{e.g.,} }

\newcommand{\sgxmonitor}{\textsf{{SgxMonitor}}}

\begin{document}

\title{Designing a Provenance Analysis for SGX Enclaves}

\author{Flavio Toffalini}
\affiliation{%
    \institution{EPFL}
    \city{Lausanne}
    \country{Switzerland}
}

\author{Mathias Payer}
\affiliation{%
    \institution{EPFL}
    \city{Lausanne}
    \country{Switzerland}
}

\author{Jianying Zhou }
\affiliation{%
    \institution{SUTD}
    \city{Singapore}
    \country{Singapore}
}

\author{Lorenzo Cavallaro}
\affiliation{%
    \institution{UCL}
    \city{London}
    \country{United Kingdom}
}



\begin{abstract}
   Intel SGX enables memory isolation and static integrity verification
   of code and data stored in user-space memory regions called enclaves. SGX
   effectively shields the execution of enclaves from the
   underlying untrusted OS. Attackers cannot tamper nor examine enclaves'
   content. However, these properties equally challenge defenders as they are
   precluded from any provenance analysis to infer intrusions inside SGX
   enclaves.
%

   In this work, we propose \sgxmonitor{}, a novel provenance analysis to
   monitor and identify anomalous executions of enclave code. To this
   end, we design a technique to extract contextual runtime information from
   an enclave and propose a novel model to represent enclaves' intrusions.
   Our experiments show that not only \sgxmonitor{}
   incurs an overhead comparable to traditional provenance tools, but it also
   exhibits macro-benchmarks' overheads and slowdowns that marginally affect
   real use cases deployment.
   Our evaluation shows \sgxmonitor{} successfully identifies enclave
   intrusions carried out by state of the art attacks while reporting no
   false positives and negatives during normal enclaves executions, thus
   supporting the use of \sgxmonitor{} in realistic scenarios.
\end{abstract}



\maketitle

\input{introduction}
\input{background}
\input{threat_model}
\input{design}
\input{model}

\input{implementation}
\input{evaluation}

\input{relatedworks}
\input{conclusion}

\bibliographystyle{ACM-Reference-Format}
\bibliography{biblio}


\appendix
\input{appendix}

\end{document}

%% file: introduction.tex
\section{Introduction}

Intel Software Guard eXtension (SGX) is an an ISA abstraction that allows
developers to define \emph{enclaves}~\cite{rozas2013intel,intel2}, small
user-space regions with strong security properties. SGX provides memory
isolation of enclaves from the underlying untrusted OS, and a remote
attestation mechanism, so-called SGX RA, to verify their integrity.
Although enclaves may host arbitrary programs, they are primarily
aimed at protecting software components that carry out specific
security- and privacy-sensitive
tasks~\cite{carefulpacking,glamdring,shinde2020besfs,244028,251582}.
Both academic~\cite{stealthdb} and
industry~\cite{signal,dashline,sharemind,fortanix,oasislabs} proposals embrace
SGX to execute such sensitive components.

In a nutshell, SGX guarantees that an enclave is properly loaded in
memory, while SGX Remote Attestation (RA) allows a remote entity to verify the
correct enclave initialization, similar to a pre-boot TPM static
code measurement. Attestation of arbitrary enclave state (\eg during or after
requests) is, so far, out of scope. As such, SGX alone has no mechanisms to
guarantee the correct runtime execution of enclaves, which remain vulnerable
against confused deputy attacks aimed at causing deviations from enclaves'
expected legitimate behaviors and lead to data
leakage~\cite{tale-two-worlds,251582,biondo2018guard,lee2017hacking,snakegx}.


Although one can equip enclaves with mechanisms tailored at counteracting
specific threats (\eg CFI or shadow stacks), these solutions simply stop an
attack without providing the analyst information about the intrusion.
In real scenarios, however, solely blocking an intrusion does not prevent
further attempts in similar contexts.
Moreover, recent works highlighted the difficulties of removing all
vulnerabilities from SGX enclaves by design~\cite{251582}.
In this regard, having insights about the attack vector thus becomes crucial
for helping analysts and engineers to improve the defenses.
This problem grows in importance when paired with memory isolated
environments, such as SGX, that shield the inspection of enclaves
a-priori~\cite{sgxforensic}.
In normal scenarios, such as standard applications or OSs,
one can extract forensic evidence of an intrusion by employing provenance
analyses~\cite{han_unicorn_2020,pasquier_runtime_2018,irshad_trace_2021,noauthor_alchemist_nodate},
that allow one to inspect the adversary movements from a stream of events
(\eg system logs, syscall invocation).
Unfortunately, the SGX memory isolation hinders provenance analysis by
disallowing any mechanism to monitor enclave executions (\eg Intel
PT~\cite{kleen2015intel} or Intel LBR~\cite{7924286,9051250}).

Observing the lack of provenance techniques for SGX, we introduce
\sgxmonitor{}, which allows an external (and legitimate) entity to inspect an
enclave runtime state and retrieve evidence of intrusion.
Having a robust provenance analysis for SGX enclaves requires us to overcome
two challenges: first, design a secure tracing mechanism for SGX
enclaves, and second, propose a model to represent useful intrusion information.
For the first challenge, we combine a lightweight enclave instrumentation
with a novel communication protocol that allows the emission of contextual
runtime information in the presence of a compromised OS, thus adhering to the
standard SGX threat model.
Our tracing is designed to offer a similar granularity as Intel PT but for SGX
enclaves, forming the foundation for provenance analyses.
Addressing the second challenge, we detect intrusion through a
novel Finite-State Machine (FSM) that extends the current models used in
SGX~\cite{costan2016intel}.
We then rely on a combination of symbolic execution and a flow-, path-, and
context-insensitive static analysis to create a FSM of the code in an enclave.
Intuitively, an enclave deviating from its FSM gives insights about the
attack vector.

To support our claims, we evaluate the properties of \sgxmonitor{} in terms of
security guarantees and usability.
To assess the security properties of \sgxmonitor{}, we test it against
SnakeGX~\cite{snakegx}, a novel data-only malware for SGX
enclaves, and specifically-crafted security benchmarks (Section
\ref{sssec:execution-flow-attacks}).
Moreover, we discuss if our communication protocol may introduce information
leakage and outline mitigation (sections~\ref{sssec:non-control-data}
and~\ref{sssec:info-leakage}).
Finally, we provide a security analysis of \sgxmonitor{}
(Section~\ref{ssec:security-analysis}).
To assess whether \sgxmonitor{} is usable in practice, we deploy it across
five use cases (Section~\ref{ssec:usage-evaluation}):
\begin{enumerate*}[label=(\roman*)]
	\item Signal Contact Discovery Service~\cite{signalrepo}
	(\textsf{Contact}), a privacy-preserving service that finds new
	contacts in
	the Signal app~\cite{signalapp};
	\item \textsf{libdvdcss}~\cite{libdvdcss}, a portable DRM library used
	by
	the VLC media player~\cite{videolan};
	\item \textsf{StealthDB}~\cite{stealthdb}, a plugin for
	PostgreSQL~\cite{momjian2001postgresql} that relies on SGX;
	\item \textsf{SGX-Biniax2}~\cite{bauman2016case}, a video game
	ported to SGX; and
	\item a \textsf{unit-test} specifically designed to stress specific
	enclave behaviors not covered by the other use cases (\ie exception
	handling).
\end{enumerate*}

In summary, we make the following contributions:
\begin{itemize}
	\item We propose \sgxmonitor{}, a novel provenance analysis system designed
	for SGX enclaves that provides:
	\begin{enumerate*}[label=(\roman*)]
		\item a new design for tracing the enclaves runtime behavior
		in the presence of an adversarial \emph{host} without relying on
		additional hardware isolation (Section~\ref{sec:system-design});
		\item a stateful representation of the SGX enclaves runtime
		properties (Section~\ref{sec:model}).
	\end{enumerate*}
	\item We assess the security properties of \sgxmonitor{} against SnakeGX
	and a specifically-crafted security benchmarks
	(Section~\ref{sssec:execution-flow-attacks}).
	Moreover, we discuss possible information leakage and propose mitigation
	(sections~\ref{sssec:non-control-data} and~\ref{sssec:info-leakage}).
    Finally, we illustrate a security analysis
    (Section~\ref{ssec:security-analysis}).
	\item We likewise evaluate the usability of \sgxmonitor{}, in particular:
	\begin{enumerate*}[label=(\roman*)]
		\item the micro-benchmark shows a median overhead of $3.9$x
		(Section~\ref{ssec:microbenchmark});
		\item the provenance analysis speed of \sgxmonitor{} is in
		line with state-of-the-art works (a median of $260$K \emph{actions}/s)
		(Section~\ref{ssec:attesation-speed});
		\item the deployment of \sgxmonitor{} does not affect the final user
		experience (\eg we smoothly played a DVD on VLC and a video game, and
		measured an average $1.6$x slowdown on PostgreSQL)
		(Section~\ref{sssec:macro-benchmar});
		\item we show a $96\%$ enclave coverage with \emph{zero} false positive, and
		investigate the trade-off between symbolic execution and static
		insensitive analysis (Section~\ref{sssec:coverage}).
	\end{enumerate*}
\end{itemize}


%% file: background.tex
\section{SGX Background}
\label{sec:background}



\emph{Enclaves} stand at the base of the SGX programming pattern.
They are contiguous memory regions that contain critical pieces of 
software and data (\eg cryptographic keys).
The isolation of SGX enclaves is handled at microcode level and is independent 
of the Operating System (OS) which is considered malicious.

SGX specifies new opcodes to interact with \emph{enclaves}.
For our work, we consider three of them:
\begin{enumerate*}[label=(\roman*)]
	\item \texttt{EENTER}, to trigger the enclave execution;
	\item \texttt{EEXIT}, to leave the enclave execution; and
	\item \texttt{ERESUME}, to resume the enclave execution after an exception.
\end{enumerate*}
Moreover, SGX uses Asynchronously Enclave Exit (\texttt{AEX}) to handle runtime
exceptions.

On top of the former opcodes, Intel provides a Software Development Kit (Intel 
SGX SDK) that organizes the enclave code as \emph{secure functions}. A process
can interact with an enclave by means of simple primitives: ECALL, to invoke a 
\emph{secure function}; ERET, to return the execution from a 
\emph{secure function}; OCALL, to invoke a function outside the 
enclave (\ie \emph{outside function}); and ORET, to resume a 
\emph{secure function} execution from an  \emph{outside function}.
In addition, the Intel SGX SDK defines dedicated \emph{secure functions} to 
handle exceptions.
The security guarantees provided by SGX ensure a strong protection against 
direct memory manipulations.
However, such protections do not hold against memory corruption vulnerabilities 
that lead to code-reuse attacks.



In addition to memory isolation, SGX introduces a Remote Attestation protocol (SGX 
RA)~\cite{vill2017sgx} that allows an external entity to verify the integrity of
an enclave. 
The SGX RA relies on the isolation offered by the CPU to protect the 
cryptographic keys. 
In particular, the SGX RA guarantees two properties:
\begin{enumerate*}[label=(\roman*)]
	\item the host machine has correctly loaded the enclave in memory,
	\item a remote entity can check the identity of the enclave and the machine (\ie CPU) 
	that is loading it.
\end{enumerate*}
Therefore, the SGX RA does not capture \emph{runtime} attacks that may deviate 
the enclave execution. 
The SGX RA provides a proof of a correctly initialized enclave but does not
consider running enclaves.
\sgxmonitor{} builds on SGX RA for enclave initialization but later continuously
verifies enclave integrity during execution.

%% file: threat_model.tex
\section{Threat Model}
\label{sec:threat-model}

In this section, we describe the threat model for \sgxmonitor{}.

\textbf{Adversary Assumptions:}
In line with the SGX assumptions~\cite{rozas2013intel}, we assume the adversary
is a host, that can attack the enclave in two ways.
(i) Exploiting classic memory-corruption errors in enclave
code~\cite{10.1145/2810103.2813646,tale-two-worlds,251582} that lead to
hijacking the enclave execution path~\cite{lee2017hacking,biondo2018guard}.
(ii) Altering the enclave communication by overhearing, intercepting, and
forging packets such as the Dolev Yao attacker~\cite{dolev1983security}.
Since the enclave has no direct access to peripherals, it requires
the OS assistance to communicate with the outside world.
Therefore, a malicious OS can intercept messages reported/received by the
enclave in the attempt to induce a wrong enclave behavior.

\textbf{Enclave Assumptions:}
We assume an enclave developed for \sgxmonitor{} follows the specification
described in sections~\ref{sec:model} and~\ref{sec:system-design}.
In particular, \sgxmonitor{} requires the source code of the enclave, that will
be
instrumented at compilation time to trace runtime enclave information
(Section~\ref{sec:implementation}).

\textbf{Out-of-Scope Attacks:}
We assume the CPU is correctly implemented, thus not prone to rollback
attacks~\cite{197191}, micro-architectural
vulnerabilities~\cite{7163052,van2017telling,203183,10.1145/3133956.3134038,kocher2019spectre,van2020lvi},
cache timing attacks \cite{206170,moghimi2017cachezoom,10.1145/3065913.3065915},
and denial-of-service from the host.
We also assume enclaves with a correct exception handler
implementation~\cite{cui2021smashex}.
Such problems are considered orthogonal to \sgxmonitor{}.

%% file: design.tex
\section{\sgxmonitor{}: System Design}
\label{sec:system-design}

\begin{figure*}[t]
	\centering
	\includegraphics[width=0.8\linewidth]{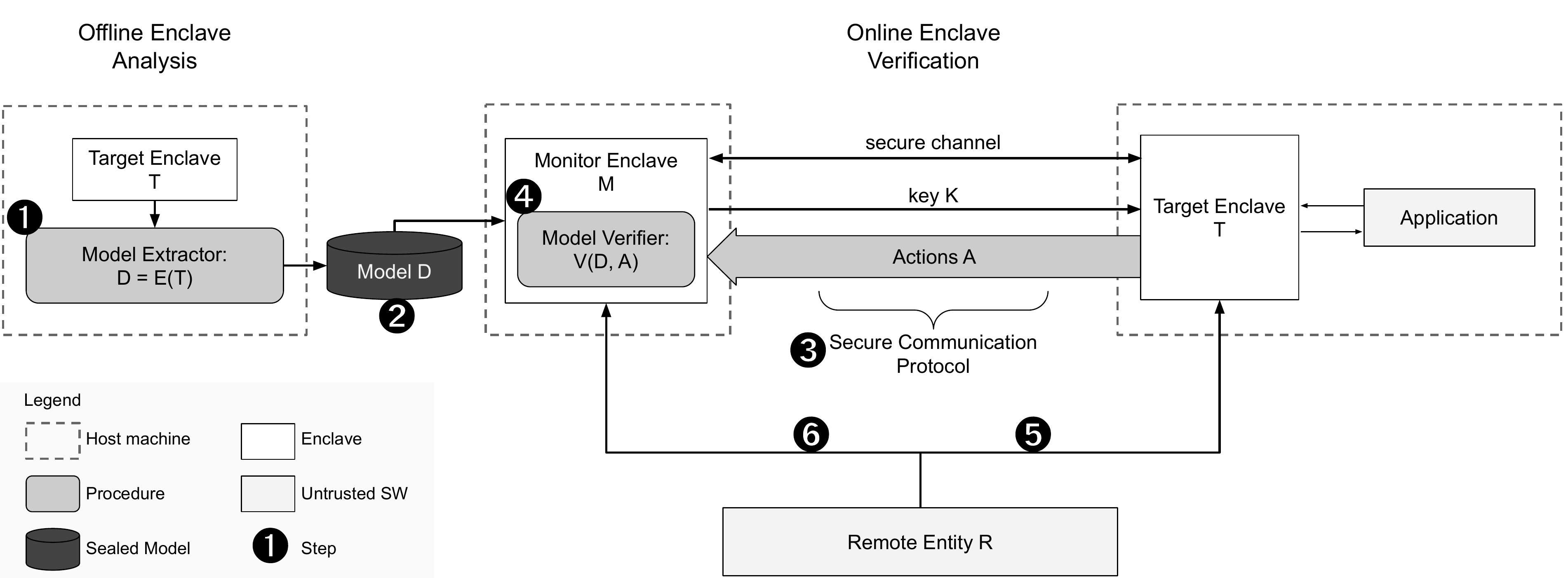}
	\caption{The \sgxmonitor{} design is composed of two distinct phases:
	Offline Enclave Analysis and Online Enclave Verification.
	During the Offline Enclave Analysis, the \emph{Module Extractor}
	analyses the \emph{target enclave} T (\circled[1]) to obtain a \emph{Model}
	D that represents the correct behavior of T (\circled[2]).
	During the Online Enclave Verification, an \emph{Application} interacts
	with T by following standard SGX mechanisms (\eg ECALL, OCALL),
	meanwhile, T sends a stream of \emph{actions} A to the \emph{monitor
	enclave} M through a \emph{secure communication protocol} (\circled[3]).
	M, then, uses a \emph{Model Verifier} to validate A against D (\circled[4]).
	Finally, a \emph{remote entity} R can perform provenance analysis of M.
	Specifically, R can verify the T static software integrity through the
	standard SGX protocol~\cite{anati2013innovative} (\circled[5]) and the
	runtime integrity of T by inquiring M about T runtime status (\circled[6]).}
	\label{fig:design}
\end{figure*}

Designing a provenance analysis that fits the SGX realm requires a re-thinking
of existing approaches.
In the original SGX threat model, the enclave content is protected against the
whole (malicious) system.
In fact, we cannot observe   enclave behavior externally (\eg through Intel
PT~\cite{kleen2015intel} or Intel LBR~\cite{7924286,9051250}).
Considering this limitation, we propose a pure software design that
allows an enclave to securely stream runtime fine-grain information, namely
\emph{actions}, similarly to Intel PT allowing an outside monitor to track
execution provenance inside the enclave without observing the computed data
flow.

This section will mainly focus on the system design by providing an
overview in Section~\ref{ssec:overview}, discuss the \emph{action} emission
mechanism in Section~\ref{ssec:code-instrumentation}, and explaining the
communication protocol in Section~\ref{ssec:secure-communication-protocol}.
The points strictly related to the model and the provenance analysis will be
detailed in Section~\ref{sec:model}.

\subsection{Overview}
\label{ssec:overview}
Figure~\ref{fig:design} illustrates the \sgxmonitor{} design, that involves
seven actors:
\begin{itemize}
	\item a \emph{target enclave} T, the enclave to monitor against attacks
	under the threat model described in Section~\ref{sec:threat-model}.
	\item a \emph{monitor enclave} M, that receives the \emph{actions} A
	generated by T.
	\item an \emph{Application}, that interacts with T through standard SGX
	specifications (\eg ECALL, OCALL),
	\item the \emph{Model} D, that represents the correct behavior of T.
	\item the \emph{Model Extractor}, that generates a model containing
	the correct behavior of T.
	\item the \emph{Model Verifier}, that validates the runtime status of T
	according to A and D.
	\item a \emph{remote entity} R, that attempts to validate both software and
	runtime integrity of T.
\end{itemize}

\paragraph{Goal and Assumptions.}
Our system lets the monitor M securely collect
\emph{actions} A from the enclave T, and later allow a remote entity R to
verify the state of T through M.
We assume T, or its host, may be compromised. Moreover, we move M into a
separate host to limit the effects of attacks against M.
Finally, R represents a system administrator that desires to
validate the integrity of T, we ensure the trustworthiness of R by employing
the standard SGX RA~\cite{anati2013innovative}.

Our goal partially overlaps with runtime remote attestation
works~\cite{scarr,abera2016c}, in which T and M are merged into a single entity.
However, such a solution does not fit our requirements because SGX enclaves
cannot be internally segmented (i.e., an enclave forms a single inseparable fault domain).
Therefore, in case of intrusion, we cannot ensure T is following the intended
design, \eg the adversary may alter the \emph{action} emission or leak
communication keys.
Conversely, uncoupling T and M (and moving
the latter into a separate host) raises the bar for attacks against T.


Overall, the design of \sgxmonitor{} is split into two distinct phases:
\emph{Offline Enclave Analysis}, and \emph{Online Enclave Verification}.
During the \emph{Offline Enclave Analysis}, the \emph{Model Extractor}
generates the \emph{Model} D representing the correct behavior of the
\emph{target enclave} T (\circled[1]).
Then, we seal D to prevent a malicious host to tamper with it (\circled[2]).
During the \emph{Online Enclave Verification}, we assume that M and T are
correctly loaded in the respective hosts.
Once T is loaded, it establishes a \emph{secure communication channel} with M by
using the standard SGX RA~\cite{anati2013innovative}, as
described in Section~\ref{ssec:secure-communication-protocol} (\circled[3]).
This channel allows T to send a stream of \emph{actions} A to M,
while an \emph{Application} can interact with T by following standard SGX
mechanisms (\eg ECALL, OCALL).
Finally, M uses the \emph{Model Verifier} to validate the runtime integrity
of T by controlling A against D (\circled[4]).
The \emph{Model Extractor} (\circled[1]) and \emph{Verifier} (\circled[4]),
along with further model details, are described in
sections~\ref{ssec:model-exctraction} and~\ref{ssec:model-validation},
respectively.

Once M correctly receives A from T, a \emph{remote entity} R can attest the
software integrity of T through the standard SGX RA~\cite{anati2013innovative}
(see Section~\ref{sec:background}). This ensures that the software
in T has been loaded properly and is not tampered with (\circled[5]).
Since we employ the standard SGX RA, we do not provide further details.
Finally, R can inquiry M regarding the runtime state of T, \ie if T still
follows the model D and, in case, where the model diverges and how
(\circled[6]).

\subsection{Action Reporting Mechanism}
\label{ssec:code-instrumentation}

T relies on an \emph{action} reporting mechanism that is resilient against the
threat model described in Section~\ref{sec:threat-model}: an intrusion inside T
(\eg exploiting a T internal error), and a malicious host.

We design the \emph{action} reporting as a dedicated function, called
\texttt{trace()}, that is included in crucial code locations of T at
compilation time.
Without loss of generality, we say all the \emph{actions} are reported through
\texttt{trace()} over a secure channel between M and T
(Section~\ref{ssec:secure-communication-protocol}).
This section mainly focuses on the reporting mechanism, while a complete
description of \emph{actions} is presented in
Section~\ref{ssec:actions-definition}.
Finally, we assume \texttt{trace()} is free from errors and an adversary
cannot exploit it to take control of T. This is reasonable since
\texttt{trace()} has a minimal implementation tailored for \emph{action}
reporting.

The intuition of our mechanism is to report an \emph{action} \emph{before} a
critical control-flow location is traversed (\ie a return instruction).
We exemplify this mechanism in Figure~\ref{fig:code-instrumentation}, in which
the program traces an \emph{action} representing a return edge to the caller
(line~\ref{fig:code-instr-ret}).
In this scenario, an adversary could attempt an intrusion by injecting a ROP
chain, report arbitrary actions, and finally hiding her presence in T.
In this case, T will report an \emph{action} representing the anomalous return
address (\ie the first ROP gadget) right before the payload is executed, thereby
producing evidence of the intrusion.
We can generalize this approach such that T reports every \emph{action}
\emph{before} they are actually executed, \ie before an intrusion begins.
We paired this mechanism with the secure communication protocol
(Section~\ref{ssec:secure-communication-protocol}) that avoids forging and
tampering with already reported \emph{actions}.
Therefore, an adversary cannot hijack T without reporting evidence about the
attack.

Our solution is robust against attempts of overwriting \texttt{trace()}.
In this case, we use the standard SGX security properties and distinguish two
cases.
First, in SGX 1.0~\cite{intel-developer-guide}, the host cannot arbitrary
alter the page permission of an enclave, this blocks any overwrite attempts by
design.
Second, for SGX 2.0, a host can change the enclave memory layout (\ie change
page permission) only upon an enclave request. However, for this to happen
an adversary has to first complete an intrusion in T, thus reporting evidence
of
the attack similarly to the previous scenario.

We thus claim the \emph{action} emission, when paired with the secure
communication protocol (Section~\ref{ssec:secure-communication-protocol}),
provides the base for our resilient provenance analysis.
We further investigate adversarial scenarios through a dedicated security
analysis im Section~\ref{ssec:security-analysis}.

\begin{figure}[t]
	\centering
	\begin{lstlisting}[style=CStyle,escapechar=|]
int fun(int a) {
  /* function body */

  // trace the indirect jump to the caller
  trace(__builtin_return_address(0));  |\label{fig:code-instr-ret}|
  return 0; |\label{fig:code-instr-retret}|
}
	\end{lstlisting}
	\caption{Example of code instrumentation. We report the action before
	critical program edges are traversed. This disallow an adversary to hijack
	T without reporting an \emph{action}. The secure protocol then ensure the
	adversary cannot forge an \emph{action}
	(Section~\ref{ssec:secure-communication-protocol}).}
	\label{fig:code-instrumentation}
\end{figure}

\subsection{Secure Communication Protocol}
\label{ssec:secure-communication-protocol}


T and M exchange messages relying on a secure communication channel resilient
against an adversarial host that may alter, eavesdrop, or forge the packets.

\paragraph{Protocol properties}
Our protocol ensures two properties:
\begin{enumerate*}[label=(\roman*)]
	\item the host cannot tamper with the packets reported by T;
	\item an adversary cannot alter or forge the packets already reported even
	if she takes control of T.
\end{enumerate*}
Note that we accept an adversary that performs a denial-of-service between T
and M.
In this case, M considers T as untrusted after a timeout.

\paragraph{Workflow}
The channel requires three steps to be established (\circled[3] in
Figure~\ref{fig:design}):
\begin{enumerate*}[label=(\roman*)]
	\item T issues a standard SGX RA~\cite{anati2013innovative} with M, thus
	ensuring a respective identity verification;
	\item M sends a secure \emph{key} K to T; and
	\item T sends the \emph{actions} to M.
\end{enumerate*}
The secure channel is shared among the threads of T, that refer to the same
key K.
We also include a thread ID into the exchanged packets, this allows M and T to
multiplex and demultiplex the communication.
The adoption of a shared key K avoids an adversary to use
the technique discussed in Dark-ROP~\cite{lee2017hacking}, we provide more
details in the Section~\ref{ssec:security-analysis}.

The validation of the transmitted \emph{actions} relies on two algorithms,
\texttt{reportLog()} and \texttt{verifyLog()}, that are illustrated in the
algorithms~\ref{alg:report-log} and~\ref{alg:verify-log}, respectively.
Both \texttt{reportLog()} and \texttt{verifyLog()} use a lock to avoid
concurrency problems.
K has the same size of the packets transmitted, thus avoiding
crypto-analysis~\cite{horstmeyer2013physical}.
Finally, we assume \texttt{reportLog()}, \texttt{verifyLog()}, and the other
supporting functions do not contain implementation errors. We consider this
reasonable since these functions are specialized for this task.

T reports a new action $A$ through instrumented code (described in
Section~\ref{ssec:code-instrumentation}).
$A$ is given as an input to \texttt{reportLog()} that encrypts and transfers it
to M over an insecure channel.
First, \texttt{reportLog()} creates a \emph{mac} by using an hash function $H_1$
and the concatenation of $A$ and the key K (Line~\ref{alg:report-log:1}
Alg.~\ref{alg:report-log}).
Then, it generates $C$ by \emph{xor}-ing the concatenation of \emph{action} $A$
and \emph{mac} with the key K (Line~\ref{alg:report-log:2}
Alg.~\ref{alg:report-log}).
At this point, it generates a new key K by hashing the
current key K with the function $H_2$ (Line~\ref{alg:report-log:3}
Alg.~\ref{alg:report-log}).
Finally, the function writes $C$ into an insecure channel
(Line~\ref{alg:report-log:4} Alg.~\ref{alg:report-log}).

On the other side, M relies on \texttt{verifyLog()} to decrypt and validate the
encrypted packets $C$.
We also assume that M receives the packets in
order.\footnote{We assume a reliable channel like TCP as in~\cite{scarr}.}
First, M decrypts the pair $(A|\text{mac})$ by \emph{xor}-ing the packet $C$
and the key K (Line~\ref{alg:verify-log:1} Alg.~\ref{alg:verify-log}).
Then, M verifies the correctness of the packet received by independently
computing \emph{mac}$^\prime$ (Line~\ref{alg:verify-log:2}
Alg.~\ref{alg:verify-log}).
If \emph{mac} and \emph{mac}$^\prime$ does not agree, $C$ was tampered during
the transmission and M sets T as untrusted (Line~\ref{alg:verify-log:4}
Alg.~\ref{alg:verify-log}).
Otherwise, $A$ is considered correct and is processed as described in
Section~\ref{ssec:model-validation} (Line~\ref{alg:verify-log:5}
Alg.~\ref{alg:verify-log}).
Finally, M generates the next key K similarly to T
(Line~\ref{alg:verify-log:6} Alg.~\ref{alg:verify-log}).

In Section~\ref{ssec:security-analysis}, we illustrate a security
analysis of the security protocol and the \emph{action} emission mechanism
(Section~\ref{ssec:code-instrumentation}).

\begin{algorithm}[t]
	\SetAlgoLined
	\DontPrintSemicolon
	\SetKwFunction{algo}{reportLog}
	\SetKwProg{myalg}{}{}{}
	\myalg{\algo{A}}{
		$\text{mac} \gets H_1(A|K)$\; \label{alg:report-log:1}
		$C \gets (A|\text{mac}) \oplus K$\; \label{alg:report-log:2}
		$K \gets H_2(K)$\; \label{alg:report-log:3}
		$write(C)$\; \label{alg:report-log:4}
	}
	\caption{Procedure used by the \emph{target} enclave to report logs in a
		secure fashion.}
	\label{alg:report-log}
\end{algorithm}
\begin{algorithm}[t]
	\SetAlgoLined
	\DontPrintSemicolon
	\SetKwFunction{algo}{verifyLog}
	\SetKwProg{myalg}{}{}{}
	\myalg{\algo{C}}{
		$(A|\text{mac}) \gets C \oplus K$\; \label{alg:verify-log:1}
		$\text{mac}^\prime \gets H_1(A|K)$\; \label{alg:verify-log:2}
		\eIf{$\text{mac}^\prime \neq \text{mac}$} {\label{alg:verify-log:3}
			$\text{untrusted}()$\; \label{alg:verify-log:4}
		}{
			$\text{process}(A)$\; \label{alg:verify-log:5}
		}
		$K \gets H_2(K)$\; \label{alg:verify-log:6}
	}
	\caption{Algorithm used by the \emph{monitor} enclave to verify the
	logs reported through \emph{reportLog()} described in
	Algorithm~\ref{alg:report-log}.}
	\label{alg:verify-log}
\end{algorithm}

%% file: model.tex
\section{\sgxmonitor{}: the Enclave Model}
\label{sec:model}

\begin{figure}[t]
	\centering
	\includegraphics[width=0.4\textwidth]{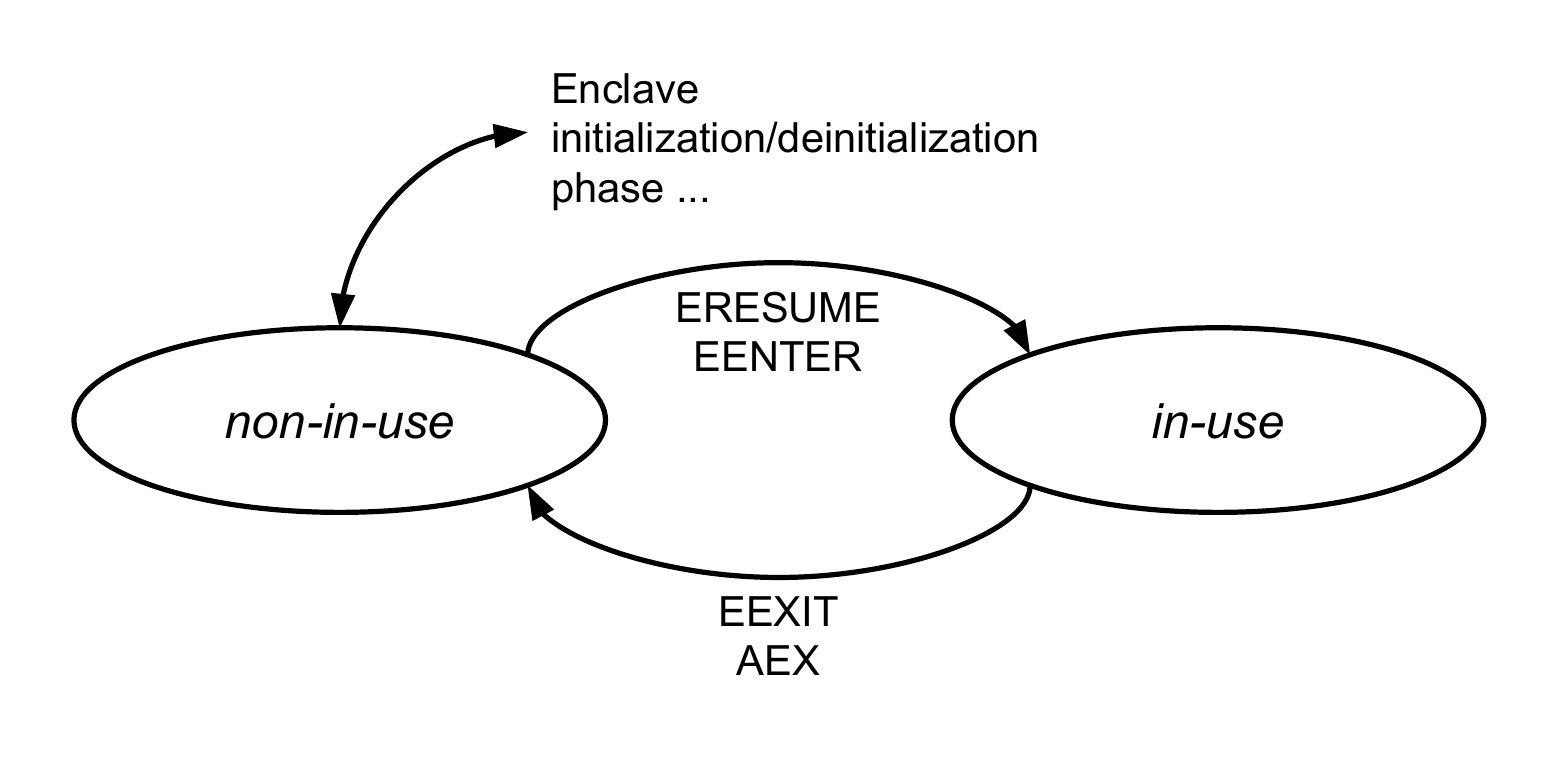}
	\caption{Standard Finite-State Machine representation of SGX
		Enclaves~\cite{costan2016intel}.}
	\label{fig:fmi-standard}
\end{figure}

To model the enclave behavior, we design a novel Finite-State Machine
that extends the standard SGX enclave life-cycle depicted in
Figure~\ref{fig:fmi-standard}.\footnote{This model is a simplified version
of~\cite{costan2016intel}.}
The standard SGX model assumes the enclave has been loaded correctly and the
host interacts with it by means of the opcodes described in
Section~\ref{sec:background}.
The model allows the enclave state to assume only two values:
\emph{non-in-use} and \emph{in-use}.
In particular, an enclave transits to \emph{in-use} state when an
\texttt{EENTER} or \texttt{ERESUME} is issued.
Then, the state returns to \emph{non-in-use} when an \texttt{EEXIT} or
\texttt{AEX} happens.
This simple model is already implemented in the microcode: the same thread
cannot enter (\ie \texttt{EENTER}) in an enclave which is
already in \emph{in-use} state; it cannot exit (\ie \texttt{EEXIT}) when the
enclave is in \emph{non-in-use}.

Intuitively, the model in Figure~\ref{fig:fmi-standard} provides limited
information about enclave health. In case of new attacks against enclaves'
code~\cite{lee2017hacking,biondo2018guard,snakegx}, we cannot
trace the enclave execution thus precluding provenance analysis in case
a-priori.


Analyzing intrusion techniques for SGX enclaves, we noticed two
patterns. Attacks either hijack the enclave execution
flow~\cite{lee2017hacking,tale-two-worlds}, or corrupt internal
enclave structures~\cite{biondo2018guard,snakegx}.
Therefore, we design the \sgxmonitor{} model to recognize those patterns.
Specifically, our model is composed by four elements:
\begin{itemize}
	\item \emph{states}, that represent the runtime values of global structures
	(Section~\ref{ssec:state}).
	\item \emph{actions}, that are meaningful binary level events (\eg
	\texttt{EENTER}, function call) (Section~\ref{ssec:actions-definition}).
	\item graphs of \emph{actions}, that are computed offline and used to
	validate runtime transactions (Section~\ref{ssec:graph-of-action}).
	\item \emph{transactions}, that are sequences of \emph{actions} leading an
	enclave from a state to the next. They express correct execution paths
	(Section~\ref{ssec:transaption-model}).
\end{itemize}
In the rest of the section, we detail state, \emph{actions}, transactions, and
graphs of \emph{actions}.
Then, we describe the \emph{Model Extractor} and \emph{Verifier} in
sections~\ref{ssec:model-exctraction} and~\ref{ssec:model-validation},
respectively


\subsection{State Definition}
\label{ssec:state}

Our model employs a state that represents important global
structures used by the Intel SGX SDK.
These structures handle operations such as \emph{outside function}
invocation and \emph{exception handling}
(Section~\ref{sec:background}) and are targeted by the
adversaries~\cite{biondo2018guard,lee2017hacking}.
Having an enclave that reaches an anomalous state provides information about
the tactic adopted for the intrusion.

Due to the multi-threading nature of enclaves, \sgxmonitor{}
traces a state for each thread~\cite{intel-developer-guide}.
The state is a triplet defined as $(usage, \\structure, operation)$.
In particular, \emph{usage} recalls the FSM meaning seen in
Figure~\ref{fig:fmi-standard} and can assume two values: \emph{in-use} and
\emph{non-in-use}.
\emph{Structure}, instead, is an hash representation of the current structure
used. If no \emph{structure} is used, it assumes \emph{null} value (\ie
$\oslash$).
Finally, \emph{operation} represents the last operation performed over the
\emph{structure}. In our model, the structures do not change over time, thus,
we trace their generation (\ie G) and consumption (\ie C).
In case no operation has been performed, we consider a \emph{null} action (\ie
$\oslash$).

In our proof of concept, we trace the generation and consumption of
\begin{enumerate*}[label=(\roman*)]
	\item \texttt{ocall\_context}, used in the \emph{outside
	functions} invocation; and
	\item \texttt{sgx\_exception\_info\_t}, used in the
	\emph{exception handing}.
\end{enumerate*}
These two structures are handled at thread granularity, thus they fit
our model.
In Appendix~\ref{sec:model-examples}, we show their FSM representation.


\subsection{Action Definition}
\label{ssec:actions-definition}

Generally speaking, an \emph{action} is a meaningful software event.
We use the \emph{actions} to represent runtime enclave transactions
(Section~\ref{ssec:transaption-model}), that allow the evolution of the
enclave state; and to build graph of \emph{actions}
(Section~\ref{ssec:graph-of-action}), that we use to validate the runtime
transactions.
In particular, we distinguish two type of \emph{actions}: \emph{generic} and
\emph{stop}.

\paragraph{Generic actions}
They identify standard software behaviors such as:
\begin{enumerate*}[label=(\roman*)]
	\item edges generated by \emph{control-flow} events; \eg \texttt{jmp},
	\texttt{call}, \texttt{ret};
	\item conditional branches (\eg \texttt{jc}); and
	\item function pointer and virtual table assignment.
\end{enumerate*}
Generic \emph{actions} do not alter the state of the enclave and they are
used to identify correct executions.
We choose these events because they are key information to represent
execution paths~\cite{scarr,hu2018enforcing,kleen2015intel,7924286,9051250}.

\paragraph{Stop actions}
They alter the state of the enclave, in particular, we consider particular SGX
opcodes and structures manipulation.
For what concerns SGX opcodes, we consider \texttt{EENTER}, \texttt{EEXIT}, and
\texttt{ERESUME}, moreover, we distinguish between \texttt{EEXIT} used for an
ERET or an OCALL, respectively. These actions alter the first field of the
state (\ie \emph{usage}): when an application enters an enclave, \emph{usage}
becomes \emph{in-use}, while \emph{usage} turns to \emph{non-in-use} when the
enclave exits.
For structures manipulations, instead, we trace whenever the enclave
generates or consumes a structure. This actions alter the \emph{structure} and
the	\emph{operation} fields in the state; \ie when an \emph{action} generates a
structure, we store the new structure hash and set \emph{operation} as G, while
we set \emph{structure} to null (\ie $\oslash$) and \emph{operation} to C when
the structure gets consumed.

Both \emph{generic} and \emph{stop actions} are formalized as a triplet:
$$
a = (type, src, value)_{cond}.
$$
In particular, \emph{type} identifies the nature of the \emph{action} (\eg
function call, \texttt{EENTER}).
\texttt{Src}, instead, is the virtual address at which the \emph{action} has
been performed.
\emph{Value} depends by the actual \emph{action} semantic; for instance; it
contains the \emph{callee} address in case of function call; a boolean value
(\ie taken or not) in case of conditional branches; a \emph{null} \emph{value}
(\ie $\oslash$) in case the \emph{action} does not require it.
Finally, \emph{cond} contains extra condition (\eg $value \ge 0$).
We provide the complete \emph{action} list in Table~\ref{tbl:actionss} grouped
by \emph{generic} and \emph{stop}.

\begin{table}[t]
	\centering
	\caption{\emph{Actions} used to define valid transactions grouped by
			\emph{generic} and \emph{stop}, respectively.}
	\begin{tabular}{ll}
		\toprule
		\multicolumn{2}{l}{\textbf{Actions}} \\ \midrule
		\multicolumn{2}{l}{\emph{Generic}} \\ \midrule
		(E, \texttt{src}|$\oslash$, \texttt{dst}|$\oslash$) & Function
		call, ind. jump, or ret inst. \\
		& \texttt{src} and \texttt{dst} can assume null value \\
		& (\ie $\oslash$) \\ 
		(B, \texttt{src}, $0|1$) & Conditional branch \\
		& ($0$: not taken, $1$: taken) \\
		(A, \texttt{src}, \texttt{addr}) & Function pointer assignment \\
		(V, \texttt{src}, \texttt{vptr}) & Virtual pointer assignment \\
		& (for C++ virtual classes) \\ \midrule
		\multicolumn{2}{l}{\emph{Stop}} \\ \midrule
		(G, \texttt{src}, \texttt{ctx}) & \texttt{ocall\_context}
		generation  \\
		(C, \texttt{src}, \texttt{ctx}) & \texttt{ocall\_context}
		consumption \\
		(J, \texttt{src}, \texttt{ctx}) &
		\texttt{sgx\_exception\_info\_t} \\
		&  generation \\
		(K, \texttt{src}, \texttt{ctx}) & \texttt{sgx\_exception\_info\_t} \\
		& consumption \\
		(N, \texttt{src}, \texttt{idx}) & \texttt{EENTER} for the \emph{secure
			function} \texttt{idx} \\
		(R, \texttt{src}, $\oslash$) & \texttt{ERESUME} \\
		(T, \texttt{src}, $\oslash$) & \texttt{EEXIT} from
		\texttt{enter\_enclave} \\
		& (ERET) \\
		(D, \texttt{src}, $\oslash$) & \texttt{EEXIT} from \texttt{do\_ocall} \\
		& (OCALL) \\
		\bottomrule
	\end{tabular}
	\label{tbl:actionss}
\end{table}

\subsection{Graphs of Actions Definition}
\label{ssec:graph-of-action}

Graphs of \emph{actions} are composed of vertexes and edges.
More precisely, vertexes and \emph{actions} are in a bijective relationship,
\ie each vertex is paired with exactly one \emph{action} and each \emph{action}
is paired with exactly one vertex.
The edges, instead, are combinations of \emph{actions} that appear at
runtime.

We opted for graphs to efficiently represent loops, that otherwise require
an unpredictable sequence of \emph{actions}.
Moreover, the graphs of \emph{actions} allow us to implement a shadow stack.
We describe the model extraction and verification in
sections~\ref{ssec:model-exctraction} and~\ref{ssec:model-validation},
respectively.

\subsection{Transaction Definition}
\label{ssec:transaption-model}

A transaction identifies a valid execution path in an enclave and
is composed of a valid sequence of \emph{actions}
(Section~\ref{ssec:actions-definition}) that makes the enclave state
evolve.
%
Formally, we indicate a transaction $P$ as following $P = [g_1, \dots, g_n, s],$
which is a sequence of \emph{generic actions} $g_i$ that terminates with a
\emph{stop action} $s$.
Intuitively, an enclave should reach a new state only through valid
transactions, otherwise we observe an anomalous enclave behavior.
We perform the transaction validation by matching the \emph{actions} received
from the monitored enclave with its graphs of \emph{actions}.
We provide the full validation algorithm in Section~\ref{ssec:model-validation}.
The combination of transactions and graph of \emph{actions} allows
one to recognize intrusion tactics~\cite{lee2017hacking,tale-two-worlds}.

\subsection{Model Extractor}
\label{ssec:model-exctraction}

\begin{algorithm}[t]
	\SetAlgoLined
	\DontPrintSemicolon
	\SetKwFunction{algo}{extractModel}
	\SetKwProg{myalg}{}{}{}
	\myalg{\algo{T}}{
		$m \gets \emptyset$\;
		\For{$f \in T.\text{instr\_functions}$}{ \label{alg:model-extractor-for}
			setSymbolicGlobalVars($T$)\; \label{alg:model-extractor-glb}
			loopAnalysis($f$)\; \label{alg:model-extractor-loop}
			setSymbolicFreeArgs($f$)\; \label{alg:model-extractor-free}
			$r \gets \text{symbolicExploration}(f)$\;
			\label{alg:model-extractor-symex}
			\If{$r.\text{isTimeout}()$} {
				$r \gets \text{insensitiveAnalysis}(f)$\;
				\label{alg:model-extractor-ins}
			}
			$m \gets m \cup (f, r.\text{graph\_of\_action})$\;
			\label{alg:model-extractor-mod}
		}
		\Return $m$
	}
	\caption{Extracting model algorithm, it takes as input the target enclave
		and returns the relative model.}
	\label{alg:model-extractor}
\end{algorithm}

The goal of the \emph{Model Extractor} (\circled[1] in Figure~\ref{fig:design})
is to automatically infer the
behavior for a given enclave. A naive approach would use a symbolic
execution~\cite{king1976symbolic} over the entire enclave.
However, this strategy does not scale to the whole code base.
Another approach would use insensitive static
analysis~\cite{coppa2017rethinking} to extract the control-flow graphs of each
function. However, this approach introduces impossible paths that increases the
attacker surface.
In our scenario, we assume that the code in an enclave implements
straight-forward functionality, such as a software daemon that implements
different features~\cite{abadi2009control} and not arbitrarily complex like,
e.g., a web-browser.
An enclave contains a relative small number of indirect call and its
software base is given.
Therefore, we take inspiration from previous compositional
analysis~\cite{calcagno2009compositional} that treats individual functions
separately.
More precisely, we extract a model for each function of the
enclave with a combination of symbolic executions and insensitive static
analysis.
We detail the model extracted in Section~\ref{sec:model} and we describe the
metrics used to define \emph{simple} programs in Section~\ref{sssec:coverage}.

The \emph{Model Extractor} takes as input a \emph{target
	enclave} T which has been instrumented at compilation time
(Section~\ref{sec:implementation}); \ie it contains extra code that traces
runtime enclave information, namely \emph{actions}; and outputs a graph of
\emph{action} for each traced function in the enclave.
T is compiled without debug information, we solely rely on global symbols to
identify the functions entry point and the global variables.
The global symbols do not contribute to the enclave measurement, thus we strip
them out after extracting the model~\cite{intel2}
(Section~\ref{sec:background}).

Overall, the extraction algorithm is described in
Algorithm~\ref{alg:model-extractor}. Given an instrumented \emph{target
	enclave} T, we analyze each instrumented function separately
(Alg.~\ref{alg:model-extractor} line~\ref{alg:model-extractor-for}).
We describe each point of the analysis in the rest of the section, while we
formalize the model in Section~\ref{sec:model}.

\textbf{Symbolic Global Variables (Alg.~\ref{alg:model-extractor}
	line~\ref{alg:model-extractor-glb}):}
Global variables might contain default concrete values that affect the symbolic
exploration.
We mitigate this issues by setting all the global variables as unconstrained
symbolic objects.
We repeat this operation for each function to clean the symbolic constraints
previously set.

\textbf{Loop Analysis (Alg.~\ref{alg:model-extractor}
	line~\ref{alg:model-extractor-loop}):}
Unbounded loops can lead to infinite symbolic
explorations~\cite{10.1007/978-3-642-36742-7_47}.
Since we are interested to reduce false positive alarms, we employed a
postdominator tree~\cite{dominators} over the static control-flow-graph to
identify the loops header in each function.
This approach is conservative and allows us to explore more execution paths,
which is our main goal.
We set the maximum to three loop iterations, similarly to previous
works~\cite{wang2009intscope}.
Our experiments show that we reach a good coverage while keeping low false
positive.

\textbf{Free Arguments Inferring (Alg.~\ref{alg:model-extractor}
	line~\ref{alg:model-extractor-free}):}
Some function requires pointers as arguments (\eg structures, objects, array),
however, current symbolic explorations do not fully handle symbolic
pointers, that might lead to a wrong or incomplete
exploration~\cite{coppa2017rethinking}.
Since we are interested to reduce false positive alarms, we opted for a
conservative approach based on static backward slicing~\cite{slicing}
to identify pointers passed as function arguments.
For each free pointer, we build an unconstrained symbolic object to help
the exploration.
This solution allows us to achieve a good coverage in
the majority of the case, as also shown in our experiments. We also
introduce custom analysis to handle corner cases, which are though a
limited number.
Finally, we deal with functions pointers by employing a conservative function
type analysis~\cite{abadi2009control}.

\textbf{Symbolic Exploration (Alg.~\ref{alg:model-extractor}
	line~\ref{alg:model-extractor-symex}):}
We primary employ a symbolic exploration~\cite{king1976symbolic} to avoid
impossible paths that, otherwise, might increase the attacker surface.
We execute the symbolic exploration after tuning the function as previously
described.
Through the exploration, we build a graph of \emph{action} for each function.

\textbf{Insensitive Static Analysis (Alg.~\ref{alg:model-extractor}
	line~\ref{alg:model-extractor-ins}):}
Since few functions of our use case experienced a symbolic execution timeout
due to their complexity (\ie too many nested loops).
We employed a fallback approach based on an insensitive static
analysis~\cite{sarkar2007flow} in which we traverse the static
control-flow-graph of the function to build the function graph of \emph{action}.
These cases are rare and they are used only if the symbolic approach fails.
We measure the frequency of this case in our evaluation.

\textbf{Building a Model (Alg.~\ref{alg:model-extractor}
	line~\ref{alg:model-extractor-mod}):}
The final enclave model is an association between functions and their
model.
We refer to Section~\ref{sec:model} for further details.
Finally, we seal the output in the \emph{monitor enclave} host to avoid
tampering.

\subsection{Model Verifier}
\label{ssec:model-validation}

The \emph{Model Verifier} (\circled[4]~Figure~\ref{fig:design}) receives a
stream of \emph{actions} from the
\emph{target enclave} T and checks whether they adhere to the \emph{Model} D.
Every \emph{action} moves T from a state to the next one, the forward jumps are
validated directly against the \emph{Model} D, while the back jumps (\eg
\texttt{ret} instructions) are validated against a shadow stack~\cite{scarr}.
These mechanisms ensure the sequence of \emph{actions} follow a correct path.
Moreover, the \emph{Model Verifier} tracks the running state of T and
identifies when the enclave reaches a wrong state.
Failing to adhering to the model D gives insights about the intrusion tactic
used to control the enclave.

%% file: implementation.tex
\section{Implementation}
\label{sec:implementation}

We provide technical details about the \emph{Compilation Unit}, the \emph{Model
Extractor}, and the \emph{secure communication channel}.

\textbf{Compilation Unit:}
%
The \emph{Compilation Unit} takes as input the \emph{target enclave} source code
and emits the instrumented enclave T.
The instrumentation injected at compilation time is considered trusted since
SGX disallows an OS to arbitrary change the enclave's page
permission, thus avoiding code replacement~\cite{intel2}.
The unit is implemented as an LLVM pass for the version $9$ (367 LoC) and a
modified version of Clang $10$ that instruments virtual pointer assignments (15
LoC added).
In the link phase, we link T with an instrumented SGX SDK to trace specific
parts of the code, \eg
in \texttt{do\_ocall} and \texttt{asm\_oret} to handle
\texttt{ocall\_context} generation/consumption; and \texttt{enter\_enclave}
to trace the entrance/exit from the enclave.
We opted for this solution because Intel does not officially support the
compilation of the SGX SDK with Clang~\cite{sgxclang}.
We based the instrumented SGX SDK on the version $2.6$.
In this process, we also include an extra secure function that issues the
\emph{secure communication channel}, and extra checks that avoid the
interaction between T and the \emph{Application} before the channel is
established (see Section~\ref{ssec:secure-communication-protocol}).

\textbf{Model Extractor:}
%
The \emph{Model Extractor} is based on angr version $8.18$ and implements the
algorithms described in Section~\ref{ssec:model-exctraction}.
We use PyVex~\cite{shoshitaishvili2015firmalice} to navigate the static CFG of the
functions, and angr symbolic engine to extract the graphs of \emph{actions}.
The \emph{Model Extractor} is composed of $8416$ LoC in total.



\textbf{Secure Communication Channel:}
%
The communication between the \emph{target enclave} T and the \emph{monitor
enclave} M is implemented by combining a TCP connection and a switchless
mechanism~\cite{tian2018switchless}.
T writes encrypted actions (see
Section~\ref{ssec:secure-communication-protocol}) into a ring-buffer that
resides in the untrusted host.
The buffer is then flushed into a TCP socket that connects T and M.
On the M side, another ring-buffer feeds the \emph{Module Verifier}.
We employ this design to reduce context switch delays~\cite{tian2018switchless}.
For the functions \texttt{reportLog()} and \texttt{verifyLog()}, we use
the \emph{sha256} implementation provided by Intel SGX SDK.
We can improve the efficiency adopting other
secure functions such as the Intel SHA extension~\cite{gulley2013intel} or
Blake2~\cite{aumasson2013blake2}.



%% file: evaluation.tex
\section{Evaluation}
\label{sec:evaluation}

We design our evaluation following the guidelines described
in~\cite{van2019sok} to avoid benchmarking flaws.
Our evaluation revolves around two main questions:
\begin{enumerate*}[label=(\textbf{RQ\arabic*})]
	\item \emph{what} insights \sgxmonitor{} provides in a provenance analysis?
	\item can I use \sgxmonitor{} in a \emph{real scenario}?
\end{enumerate*}
We answer \textbf{R1} in Section~\ref{ssec:security-properties} by testing
the \sgxmonitor{} security guarantees against a set of modern SGX attacks.
We answer \textbf{RQ2} in Section~\ref{ssec:usage-evaluation}.
More precisely, we measure micro-benchmark
(Section~\ref{ssec:microbenchmark}), provenance analysis speed
(Section~\ref{ssec:attesation-speed}),
macro-benchmark (Section~\ref{sssec:macro-benchmar}),
and discuss the model extraction (Section~\ref{sssec:coverage}).

\subsection{RQ1 - Security Evaluation}
\label{ssec:security-properties}

We evaluate the security guarantees of \sgxmonitor{} from multiple
perspectives.
First, we demonstrate the provenance capability of \sgxmonitor{} to intercept
modern execution-flow attacks (Section~\ref{sssec:execution-flow-attacks}).
Then, we discuss non-control data attacks and discuss mitigation
(Section~\ref{sssec:non-control-data}) and analyze the impact of \sgxmonitor{}
in side-channels scenarios (Section~\ref{sssec:info-leakage}).
Finally, we provide a security analysis of the \sgxmonitor{} design
(Section~\ref{ssec:security-analysis}).

\subsubsection{Execution-flow attacks}
\label{sssec:execution-flow-attacks}

Since SGX does not allow one to arbitrary change the page permission of a
running enclave,
researchers adapted memory-corruption errors to hijack the enclave execution.
To test the properties of \sgxmonitor{} against this class of attacks, we
choose two security benchmarks: \textsf{SnakeGX}~\cite{snakegx}, which is an
enclave infector for SGX enclaves; and a security benchmark
that evaluates the correctness of the shadow stack defense.

\textbf{SnakeGX.}
This is a data-only malware designed to implant a permanent backdoor into
legitimate SGX enclaves.
\textsf{SnakeGX} is an extension of the work of Biondo et.
al~\cite{biondo2018guard} and is based on code-reuse techniques.
\textsf{SnakeGX} is composed of two phases:
\begin{enumerate*}[label=(\roman*)]
	\item an \emph{installation phase}, that uses a classic
	ROP-chain~\cite{carlini2014rop} to install the payload inside
	the \emph{target enclave}; and
	\item a \emph{backdoor activation}, that exploits a design error of the
	Intel SGX SDK to trigger the payload previously installed.
\end{enumerate*}
\textsf{SnakeGX} managed to bypass the current SGX protections.
Therefore, once installed, an external observer cannot realize the presence of
\textsf{SnakeGX} in the \emph{target enclave}.
For our evaluation, we recompiled the victim enclave including \sgxmonitor{},
and we adjusted the \emph{gadgets} addresses of \textsf{SnakeGX} accordingly.
Then, we extracted the model, execute the malware, and finally, traced
the \emph{actions} reported.
The results show that \sgxmonitor{} recognized either the
\emph{installation phase} and the \emph{backdoor activation}.
In particular, the \emph{installation} relies on a classic ROP-chain,
therefore, \sgxmonitor{} identified an unknown \emph{action} pointing a
\emph{gadget}. In this way, \sgxmonitor{} gave an information about an intrusion inside
the enclave.
The \emph{backdoor activation}, instead, restores a corrupted
\texttt{ocall\_context} (crafted during the installation). In this case,
\sgxmonitor{} observed the restoring of an anomalous state.
Notably, previous works cannot identify the error design used in this
phase~\cite{251582}, unlike \sgxmonitor{}.

To sum up, \sgxmonitor{} gave insights about an intrusion into the \emph{target
enclave} by revealing the gadget used for the \emph{installation phase} and the deviated
patch that lead to the tampered structure in the \emph{backdoor activation}.

\textbf{Shadow stack protection.}
We evaluate the shadow stack implemented in \sgxmonitor{}.
In particular, we want to identify an adversary able to overwrite the
\emph{return address} of a function with a valid location that
is, however, incoherent with the call stack.
To this end, we built a custom enclave that allows such attacks, we compiled it
with \sgxmonitor{}, extracted the model, and finally, run the attack.
The results show that \sgxmonitor{} managed to identify execution
flows incoherent with the call stack, thus pinpointing a possible local buffer
overflow and in which function it happened.

\textbf{Final Notes.}
We remark that standard mitigation deployed inside an enclave (\eg CFI or
shadow stacks) lack any insight about the attack performed.
On the contrary, \sgxmonitor{} provides fine-grain information about the
intrusion.

\subsubsection{Non-control data attacks}
\label{sssec:non-control-data}

We discuss if the communication protocol between \emph{monitor} and
\emph{target enclave} may brace the adversary capabilities in non-control data
attacks~\cite{269251,hu2015automatic}.
Before we analyze this problem, we remark that all the packets have the same
size by design, and the cryptographic key changes at any packet reported (see
Section~\ref{ssec:secure-communication-protocol}).
Therefore, an adversary can only analyze the packets timestamp.

These attacks do not hijack the execution-flow, for instance, an enclave may
contain a password checking algorithm that matches one character at time.
In this example, the number of packets suggests the number of characters
guessed, thus reducing the combination.
We can mitigate this attack with the introduction of dummy packets (from $0$ to
$k$) and adding a random dummy delay (from $0$ to $t$).
This will increase the micro-benchmark overhead of a factor $(k + t)$x in the
worst case.
However, such defenses would be applied to specific code portions (\eg
in the password checking), thus incurring a minimal overhead footprint
overall.  (The idea is similar to adding countermeasures against
timing-based attacks~\cite{10.1007/978-3-642-25385-0_26}.)

\subsubsection{Side-channels attacks}
\label{sssec:info-leakage}

We study the implication of \sgxmonitor{} in side-channel attacks.
First, we focus on crypto analysis. In this case, an adversary may use the
number of packets reported to attack the cryptographic algorithms in the
enclave. However,
modern cryptographic algorithms have been proven chosen-ciphertext attack
secure~\cite{barthe2011beyond}. Therefore, leakage of ciphertext packets does
not improve the adversary's capabilities~\cite{wee2010efficient}.
An adversary may however count the packets exchanged by the communication
protocol to analyze the enclave execution and locate likely code
positions.
We dissect this scenario in two cases.
(i) The code location could be used in \emph{execution-flow attacks},
therefore, an adversary will trigger an anomalous execution that will be
detected by \sgxmonitor{}, as we discuss in
Section~\ref{sssec:execution-flow-attacks}.
(ii) The code location could be used in \emph{non-control data
	attacks}, that we discuss in Section~\ref{sssec:non-control-data}.

\subsubsection{Security Analysis of the System Design}
\label{ssec:security-analysis}
We discuss the security properties of the \sgxmonitor{} design
(Section~\ref{sec:system-design}) with respect to our threat model
(Section~\ref{sec:threat-model}).

\paragraph{Attacks before protocol establishing}
An adversary may target T before it establishes the secure channel with M.
To mitigate this attack surface, we enforce that all the security functions of
T are disabled until T and M completely initialize the security protocol.
In particular, the \emph{Application} must invoke a dedicated secure function
of T before it may use any other secure function.
We insert additional checks that ensure no other functionality of T is active
until T and M successfully established the channel.
This design avoids an adversary to attack T before M starts monitoring it.

\paragraph{Defense against a tampered enclave T}
Our protocol resists an adversary that exploits T.
In this case, the adversary may abuse a memory corruption error to
divert the enclave execution path.
However, we instrument the code of T such that it reports the \emph{action}
before the enclave traverses the hijacked edge
(Section~\ref{ssec:code-instrumentation}).
Therefore, the \emph{action} results are already encrypted and shipped, while K has
been altered by the hash function
(Section~\ref{ssec:secure-communication-protocol}).
We face three scenarios here:
\begin{enumerate*}[label=(S\arabic*)]
    \item the compromised \emph{action} reaches M, thus M
    recognizes the attack;
    \item the host drops the \emph{action} before reaching M, thus M
    recognizes the attack after a timeout; and
    \item the adversary attempts to forge a new valid \emph{action}, however,
    she cannot retrieve K after \texttt{reportLog()} invocation (\ie a new K is
    produced).
\end{enumerate*}
In all these cases, M will observe an anomaly in the protocol or T
behavior, finally setting T as untrusted.

Sharing the same key K among the threads defeats the tactic described
in modern enclave attacks~\cite{lee2017hacking}.
In their scenario, an \emph{adversary} exploits a thread
to leak information (\ie the key K) from another thread.
In our design, leaking K forces a thread to report an \emph{action} X
representing the attack. Moreover, \texttt{reportLog()} ensures the
\emph{actions}
follows a specific order.
Therefore, either X reaches M, thus revealing the attack; or X is
dropped, thus showing an anomaly.

\noindent\begin{minipage}{\columnwidth}%
	\vspace{1mm}
	\begin{mdframed}[style=HighlightFrame]\
\textbf{Security Evaluation---Take Away.} Our evaluation shows
that \sgxmonitor{} provides useful information about the payload used in modern
state-of-the-art attacks (\ie \textsf{SnakeGX} and shadow stack protection).
Moreover, we discuss the possible information leakage and we show that,
in practice, it does not improve the adversary capabilities.
We also propose information leaking mitigation and discuss
the scenarios for which they are more suitable.
Finally, we discuss a security analysis of our system design.
	\end{mdframed}
\end{minipage}

\subsection{RQ2 - Usage Evaluation}
\label{ssec:usage-evaluation}

\begin{table*}[t]
	\centering
	\caption{Detailed information for the of five use cases used in our
		evaluation: \textsf{Contact}~\cite{signalrepo},
		\textsf{libdvdcss}~\cite{libdvdcss},
		\textsf{StealthDB}~\cite{stealthdb},
		\textsf{SGX-Biniax2}~\cite{bauman2016case}, and a \textsf{unit-test}.}
	\begin{tabular}{lrrrrrrrrrr}
		\toprule

		\multirow{2}{*}{Use case} &
		\multirow{2}{*}{LoC} &
		\# secure &
		\multicolumn{2}{c}{\emph{cycl. cmplx.}} &
		\multicolumn{2}{c}{\# nodes in CFG} &
		\multicolumn{2}{c}{\# edges in CFG} &
		\multicolumn{1}{c}{\# direct} &
		\multicolumn{1}{c}{\# indirect}
		\\

		& &
		\multicolumn{1}{c}{function} &
		\multicolumn{1}{c}{$\mu$}  &
		\multicolumn{1}{c}{$\sigma$} &
		\multicolumn{1}{c}{$\mu$}  &
		\multicolumn{1}{c}{$\sigma$} &
		\multicolumn{1}{c}{$\mu$}  &
		\multicolumn{1}{c}{$\sigma$} &
		\multicolumn{1}{c}{calls} &
		\multicolumn{1}{c}{calls}
		\\ \midrule

		\textsf{Contact}~\cite{signalrepo} & $4138$ & $6$ & $5.03$ & $5.04$ &
		$24.89$ & $22.74$ & $26.67$ & $27.82$ & $1085$ & $16$ \\
		\textsf{libdvdcss}~\cite{libdvdcss} & $3438$ & $4$ & $6.55$ & $6.07$ &
		$38.71$ & $31.28$ & $39.67$ & $37.95$ & $1084$ & $2$ \\
		\textsf{StealthDB}~\cite{stealthdb} & $10351$ & $3$ & $6.35$ & $4.72$ &
		$36.14$ & $23.38$ & $40.40$ & $27.51$ & $1203$ & $2$ \\
		\textsf{SGX-Biniax2}~\cite{bauman2016case} & $4696$ & $7$ & $3.73$ &
		$4.20$ & $18.56$ & $16.25$ & $20.19$ & $20.02$ & $583$ & $2$ \\
		\textsf{unit-test} & $583$ & $3$ & $4.06$ & $5.25$ & $18.44$ & $17.53$
		& $18.75$ & $21.95$ & $137$ & $2$ \\
		\bottomrule
	\end{tabular}
	\label{tbl:preliminary_analysis}
\end{table*}

We describe the use cases used, the experiment setup, and discuss the
impact of \sgxmonitor{} in real projects.

\paragraph*{Use Cases}
We identified $10$ open-source projects that use SGX.
Most of them do not compile because they refer to old SGX features or they are
incompatible with Clang.
Among them, we choose five ones:
\begin{enumerate*}[label=(\roman*)]
	\item \textsf{Contact}~\cite{signalrepo}, the contact discovery service
	used by Signal app~\cite{signalapp};
	\item an SGX porting of \textsf{libdvdcss}~\cite{libdvdcss}, a portable DRM
	algorithm used by VLC media player~\cite{videolan};
	\item \textsf{StealthDB}~\cite{stealthdb}, a
	PostgreSQL~\cite{momjian2001postgresql} plugin that uses SGX to encrypt
	tables;
	\item \textsf{SGX-Biniax2}~\cite{bauman2016case}, an SGX porting of the
	open-source game Biniax2~\cite{biniax2}; and
	\item a \textsf{unit-test} to validate corner cases of the enclave
	behaviors not covered previously, like
	exception handling.
\end{enumerate*}
In Table~\ref{tbl:preliminary_analysis}, we indicate the line of code (LoC) and
the number of secure functions for each use case.

We use \textsf{Contact}, \textsf{StealthDB}, \textsf{SGX-Biniax2}, and the
\textsf{unit-test} to stress micro-benchmarks
(Section~\ref{ssec:microbenchmark}) and provenance analysis speed
(Section~\ref{ssec:attesation-speed}).
We use \textsf{libdvdcss}, \textsf{StealthDB}, and \textsf{SGX-Biniax2} for
macro-benchmarks (Section~\ref{sssec:macro-benchmar}).
All the five use cases are used for model extraction analysis
(Section~\ref{sssec:coverage}).

\paragraph*{Experiment Setup}
All the experiments were performed on a Linux machine with kernel version
$4.15.0$ and equipped with an Intel i7 processor and $16$GB of memory.
We set the CPU power governor as \emph{power save}.
Moreover, we perform a warm-up round for each \emph{secure function} before
actually recording the performances.
%

\begin{figure*}[t]
	\centering
	\begin{subfigure}[t]{.47\textwidth}
		\centering
		\includegraphics[width=\linewidth]{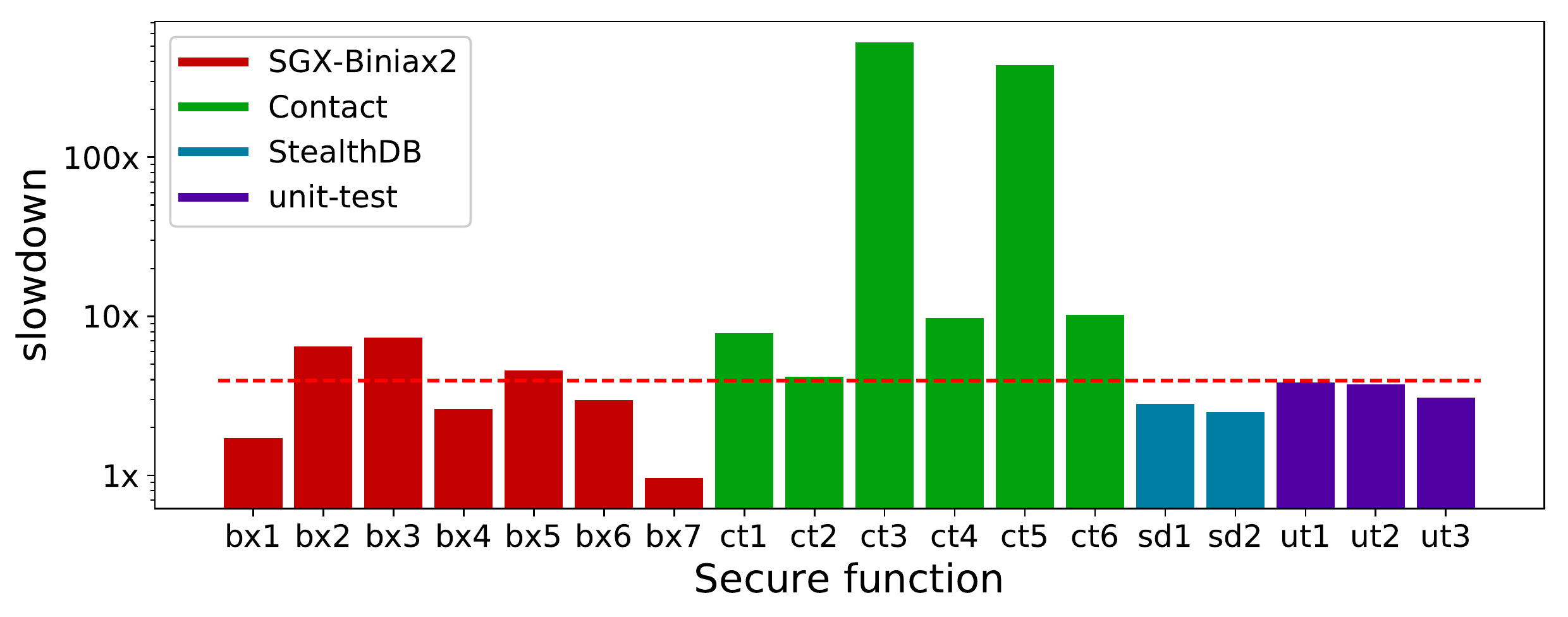}
		\caption{Overhead of vanilla secure functions versus \sgxmonitor{}
		secure
			functions of \textsf{Contact} (ct\emph{x}), \textsf{SGX-Biniax2}
			(bx\emph{x}), \textsf{StealthDB} (sd\emph{x}) and
			\textsf{unit-test} enclave (ut\emph{x}) expressed in logarithmic
			scale. Median overhead is around $3.9$x and is depicted as a dashed
			line.}
		\label{fig:multiply}
	\end{subfigure}
	\hfill
	\begin{subfigure}[t]{.47\textwidth}
		\centering
		\includegraphics[width=\linewidth]{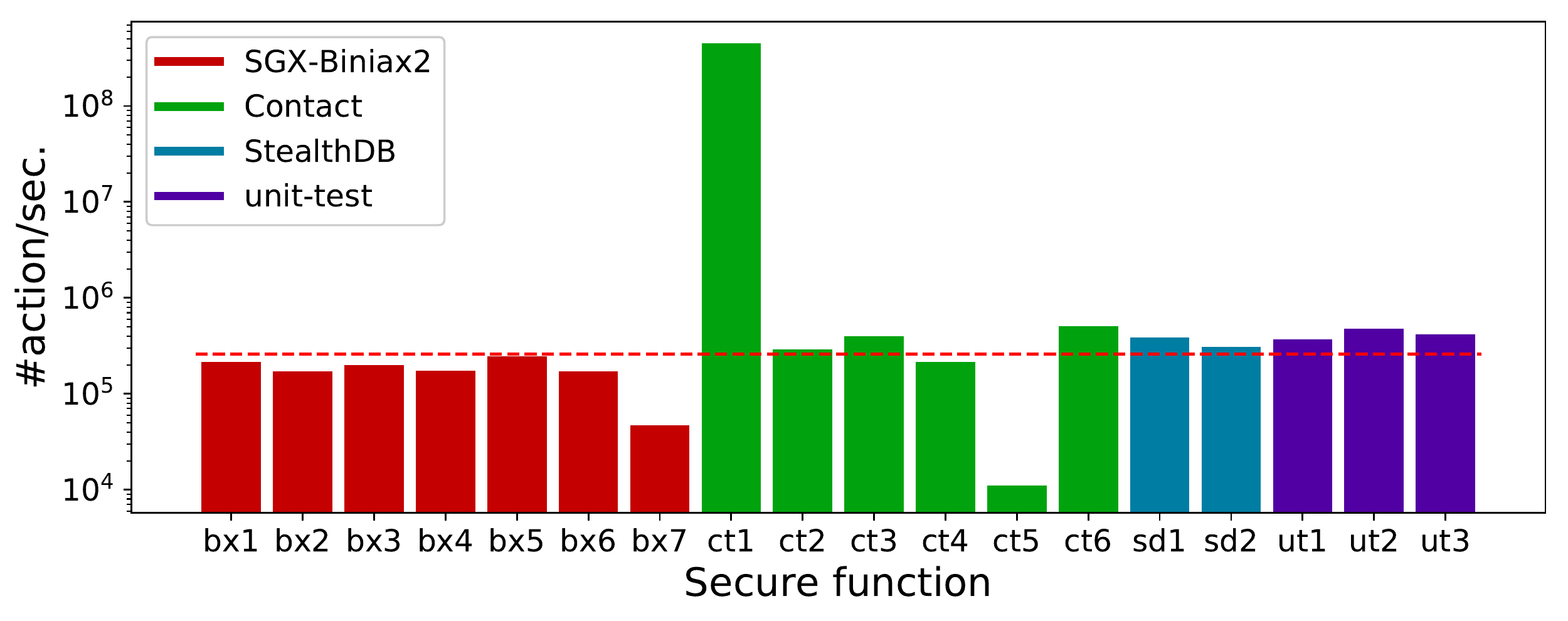}
		\caption{Number of \emph{actions} processed per second of
		\textsf{Contact} (ct\emph{x}), \textsf{SGX-Biniax2}
		(bx\emph{x}), \textsf{StealthDB} (sd\emph{x}) and
		\textsf{unit-test} enclave (ut\emph{x}). Median value is $260$K
		\emph{action} per second and is depicted as a dashed line.}
		\label{fig:action-second}
	\end{subfigure}
	\caption{\sgxmonitor{} micro-benchmark and \emph{action} speed measurement
		evaluation.}
	\label{fig:fig}
\end{figure*}

\subsubsection{Micro-benchmark}
\label{ssec:microbenchmark}

In this experiment, we measure the overhead of the single secure functions
with \sgxmonitor{} and without (\ie vanilla).
We perform this experiment on \textsf{Contact}, \textsf{SGX-Biniax2},
\textsf{StealthDB} and the \textsf{unit-test} enclave.
The results are shown in Figure~\ref{fig:multiply}.
In most of the cases, \sgxmonitor{} introduces an overhead less than or equal
to
$10$x (bx$1$-$7$, ct$1$-$2$, ct$4$, ct$6$, ut$1$-$3$) with a median overhead of
$3.9$x.
Only two secure functions show an overhead over $100$x (ct$3$ and ct$5$).

\noindent\begin{minipage}{\columnwidth}%
\vspace{3mm}
	\begin{mdframed}[style=HighlightFrame]
\textbf{Micro-benchmark---Take Away.}
A major source of overhead is incurred by the hash functions in the
secure communication protocol
(Section~\ref{ssec:secure-communication-protocol}), as observed in similar
works~\cite{scarr,abera2016c,aberadiat}.
Different hash functions can ease the overhead, \eg the Intel SHA
extension~\cite{gulley2013intel} or Blake2~\cite{aumasson2013blake2}.
However, This result does not really affect the performance of \sgxmonitor{}
that is in line with similar works~\cite{scarr} for the of analysis speed
(Section~\ref{ssec:attesation-speed}) and final user experience
(Section~\ref{sssec:macro-benchmar}).
\end{mdframed}%
\vspace{1mm}
\end{minipage}%



\subsubsection{Provenance Analysis Speed}
\label{ssec:attesation-speed}

Figure~\ref{fig:action-second} measures the provenance analysis speed in terms
of number of \emph{actions} reported and validated per second (on the y-axes)
for each \emph{secure functions} of \textsf{Contact}, \textsf{SGX-Biniax2},
\textsf{StealthDB}, and the \textsf{unit-test} enclave (on the x-axes).
The execution time encompasses the context-switch delay, \emph{actions}
emission, transmission, and verification at the \emph{monitor} side.

All the secure functions, but ct$1$, ct$5$ and bx$7$, express a throughput
that ranges from $167$K \emph{action}/sec (bx$2$) to $496$K
\emph{action}/sec (ct$6$), with a median value of $260$K \emph{action}/sec.

\noindent\begin{minipage}{\columnwidth}%
\vspace{1mm}
	\begin{mdframed}[style=HighlightFrame]
\textbf{Provenance Analysis Speed---Take Away.}
These figures are in line with the previous works~\cite{scarr}.
ct$1$, instead, reports a fewer number of  \emph{actions} and biases the
analysis
speed. Finally, bx$7$ and ct$5$ perform sealing
operations~\cite{anati2013innovative} and thus introduce an extra delay per
\emph{action}.
\end{mdframed}%
 \vspace{1mm}
\end{minipage}%


\begin{figure*}[t]
	\centering
	\begin{subfigure}[b]{0.49\textwidth}
		\centering
		\includegraphics[width=\textwidth]{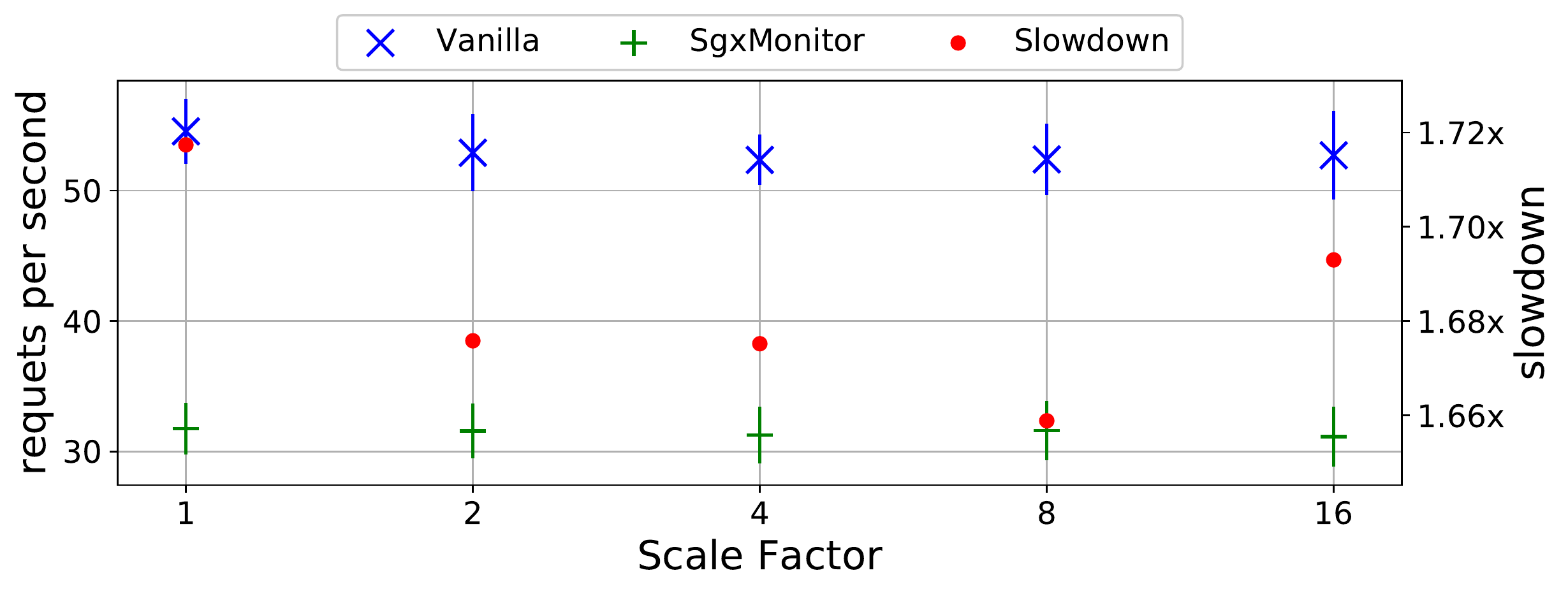}
		\caption{Overhead of \textsf{StealthDB} vanilla and with \sgxmonitor{}
		measured as requests per second.
		Overall, \sgxmonitor{} introduces an average slowdown of $1.68$x with a
		standard deviation of $0.02$x.}
		\label{fig:request_per_second}
	\end{subfigure}
	\hfill
	\begin{subfigure}[b]{0.49\textwidth}
		\centering
		\includegraphics[width=\textwidth]{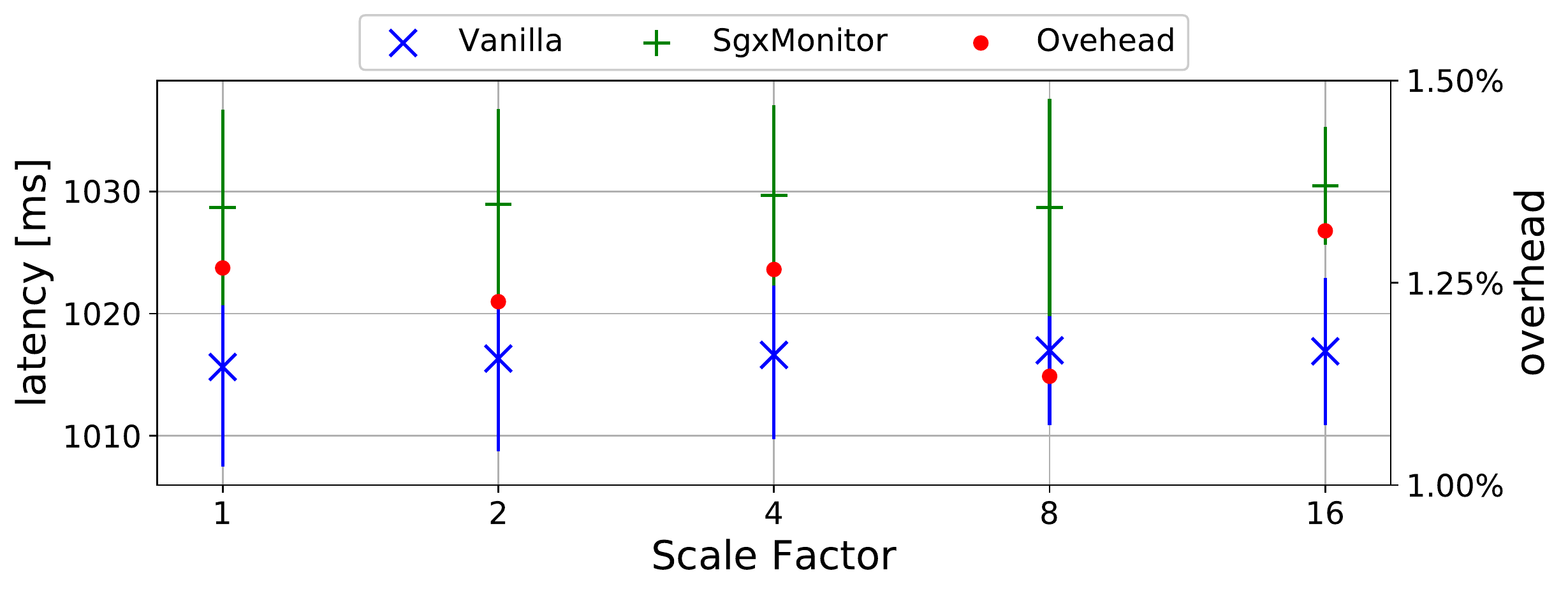}
		\caption{Overhead of \textsf{StealthDB} vanilla and with \sgxmonitor{}
		measured as latency (ms).
		Overall, \sgxmonitor{} introduces an average overhead of $1.24$\% with
		a
		standard deviation of $0.06$\%.}
		\label{fig:latency_2}
	\end{subfigure}
\caption{\textsf{StealthDB}~\cite{stealthdb} performances measured against
OLTP~\cite{oltp} benchmark and expressed as request per second and latency.
We evaluated \textsf{StealthDB} vanilla and with \sgxmonitor{}, in particular,
we
run $10$ measurements for each scale factor (from $1$ to $16$) and plot average
and standard deviation for requests per second and latency, respectively.}
\label{fig:request_per_second_latency}
\end{figure*}

\begin{figure*}[t]
	\centering
	\begin{subfigure}[t]{0.49\textwidth}
		\centering
		\includegraphics[width=\textwidth]{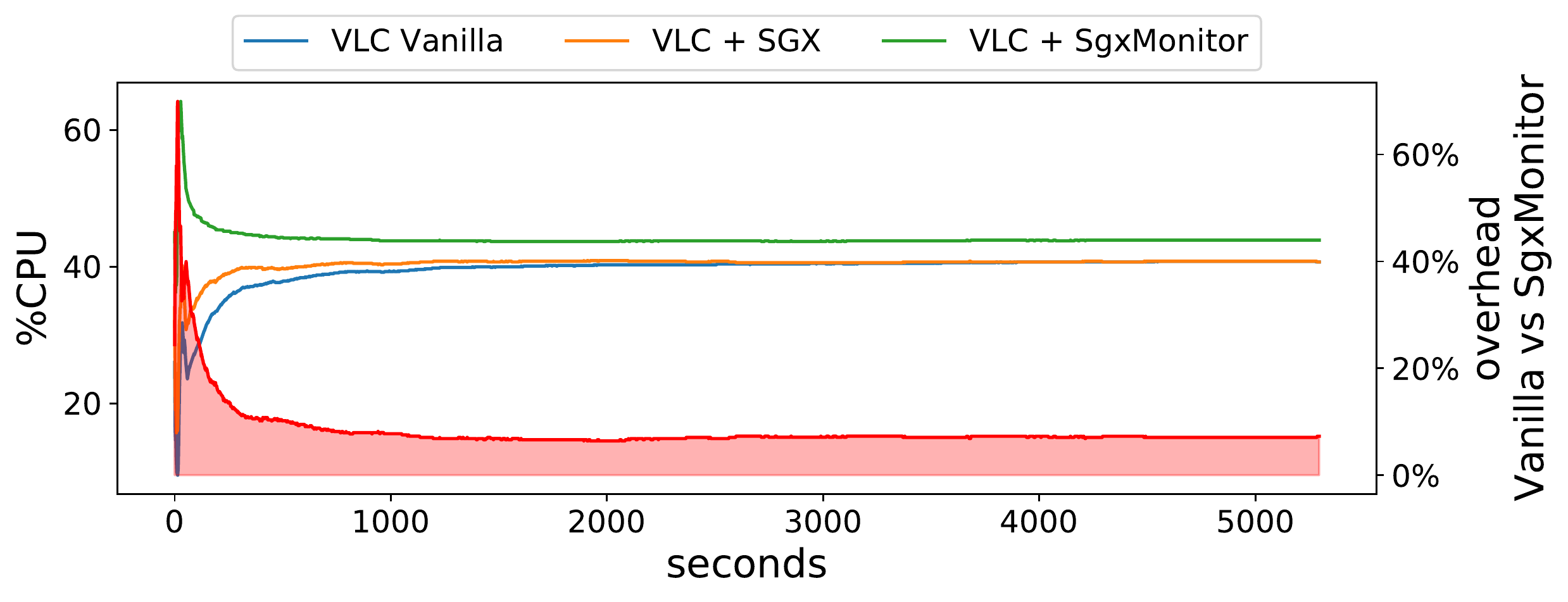}
		\caption{Overhead of VLC with \textsf{libdvdcss} vanilla, plus SGX, and
		plus \sgxmonitor{}, respectively. We measure the percentage of CPU
		usage
		while playing the same DVD with the three settings.
		After an initial adjusting phase, the overhead drops and reaches a
		plateau lower then $10$\%.}
		\label{fig:vlc_performance}
	\end{subfigure}
	\hfill
	\begin{subfigure}[t]{0.49\textwidth}
		\centering
		\includegraphics[width=\textwidth]{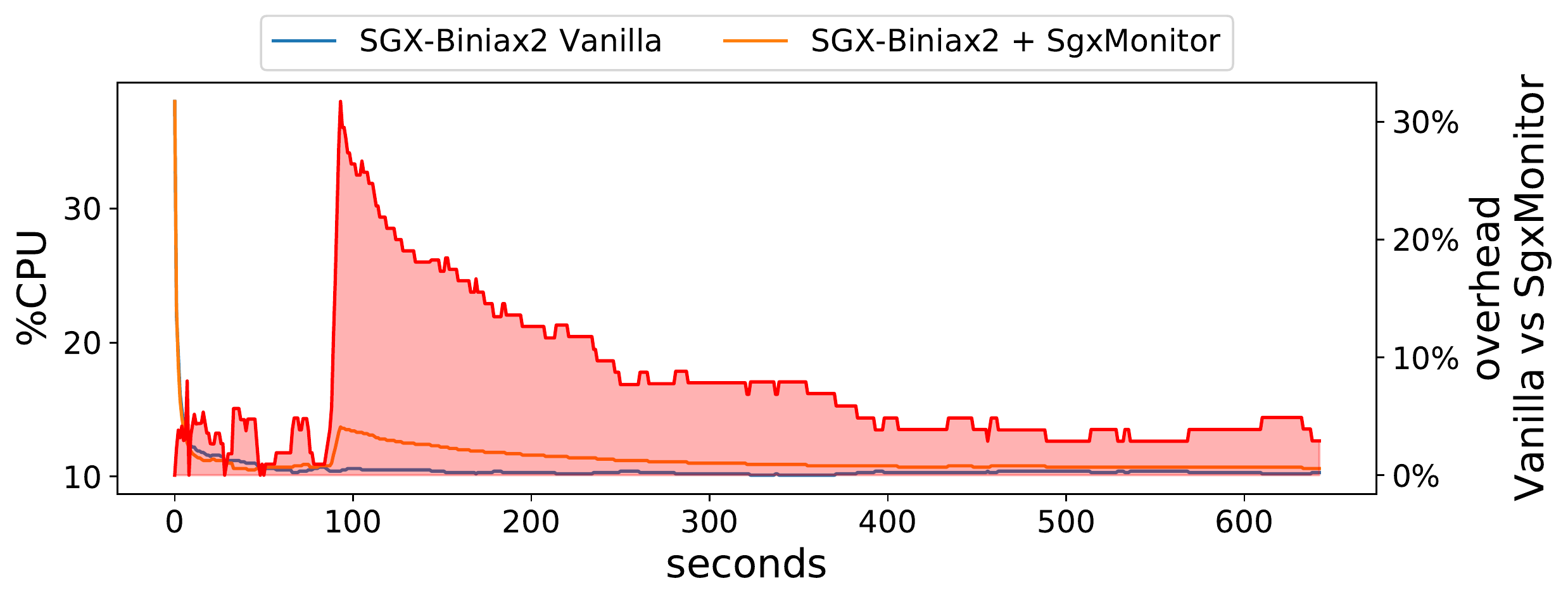}
		\caption{Overhead of \textsf{SGX-Biniax2} vanilla and with
		\sgxmonitor{},
		respectively. We measure the percentage of CPU usage while playing the
		game for the same amount of time (around $20$m).
		After an initial adjusting phase, the overhead drops and reaches a
		plateau at around $5$\%.}
		\label{fig:biniax2_performance}
	\end{subfigure}
	\caption{Macro-benchmark of \textsf{libdvdcss}~\cite{libdvdcss}, deployed
	over VLC media player~\cite{videolan}, and
	SGX-Biniax2~\cite{bauman2016case}. In both cases, we measured the CPU usage
	and the overhead introduced by \sgxmonitor{} versus the vanilla version of
	the
	software.}
	\label{fig:vlc_biniax2_performance}
\end{figure*}

\subsubsection{Macro-benchmark}
\label{sssec:macro-benchmar}

We investigate the impact of \sgxmonitor{} in three real applications.
\begin{enumerate*}[label=(A\arabic*)]
	\item \textsf{StealthDB}~\cite{stealthdb}, which is a plugin for
	PostgreSQL~\cite{momjian2001postgresql} based on SGX.
	\item \textsf{libdvdcss}~\cite{libdvdcss}, which is a DRM library used in
	VLC media player~\cite{videolan}.
	\item \textsf{SGX-Biniax2}~\cite{bauman2016case}, which is an SGX porting
	of the open-source game Biniax2~\cite{biniax2}.
\end{enumerate*}

\paragraph*{StealthDB}
We replicated the same experiments described in the
original paper~\cite{stealthdb}.
In particular, we deployed \textsf{StealthDB} over a
PostgreSQL~\cite{momjian2001postgresql} version $10.15$
and we run the database benchmarking tool OLTP~\cite{oltp} by using the
five scale factors indicated in the original work.
Then, we reported the requests per second and the latency in
figure~\ref{fig:request_per_second} and~\ref{fig:latency_2}, respectively.
For each scale factor, we run $10$ experiments and indicate average and
standard deviation.
Overall, \sgxmonitor{} introduces an average slowdown of $1.68$x and an
overhead
of $1.25\%$ in terms of requests per second and latency, respectively.

\paragraph*{libdvdcss}
We measured the CPU impact of \sgxmonitor{} over \textsf{libdvdcss}, which is
an
DRM library used in VLC media player~\cite{videolan}.
For the experiment, we used a VLC version $3.0.8$, on which we deployed three
versions of \textsf{libdvdcss}~\cite{libdvdcss}: vanilla, with SGX, and with
\sgxmonitor{}.
During the experiment, we played a DVD for around one hour and half while
sampling the CPU usage every second.
Figure~\ref{fig:vlc_performance} shows the result of our experiment, after a
first adjusting phase, the overhead reaches a plateau below $10\%$.
Furthermore, we did not experience any delay or interruption while playing the
DVD in any of the three configurations.

\paragraph*{SGX-Biniax2}.
We measured the CPU impact of \sgxmonitor{} over
\textsf{SGX-Biniax2}~\cite{bauman2016case}, an example of video game porting
that uses SGX for data protection.
In particular, we played the game for around $20$ minutes and we sampled the
CPU usage every second.
Figure~\ref{fig:biniax2_performance} shows the result of our experiment,
similarly to \textsf{libdvdcss}, we observed a first adjusting phase followed
by a plateau at around $5\%$.
Furthermore, we did not experience any delay or interruption while playing
\text{SGX-Biniax2} in any of the two configurations.

\noindent\begin{minipage}{\columnwidth}%
\vspace{2mm}
	\begin{mdframed}[style=HighlightFrame]
\textbf{Macro-benchmark---Take Away.}
Our results show that the overhead introduced by
\sgxmonitor{} is overall limited, \eg the slowdown in
\textsf{StealthDB} is lower than the micro-benchmarks (\ie $1.6$x vs $3.9$x) and
the CPU overhead expressed by \textsf{libdvdcss} and \textsf{SGX-Biniax2} shows
a limited plateau.
Therefore, we conclude that \sgxmonitor{} does not affect the final user
experience and can be included into projects that either require occasional
enclave interactions (like DRM protection) or are more computational intense
(like a database).
\end{mdframed}%
 \vspace{1mm}
\end{minipage}%


\begin{table*}[t]
	\centering

	\caption{Coverage analysis over our five use cases:
		\textsf{Contact}~\cite{signalrepo},
		\textsf{libdvdcss}~\cite{libdvdcss},
		\textsf{StealthDB}~\cite{stealthdb},
		\textsf{SGX-Biniax2}~\cite{bauman2016case}, and a unit-test.
		The	results show that the analysis covers from $91.4\%$ to $96.6\%$ of
		the \emph{actions} in around $2$ hours and $20$ minutes in total
		($8146.11$s). Furthermore, we did not observe any false positive during
		our experiments, meaning we covered a significant portion of code.
		In the right part of the table, we indicate the \emph{actions} explored
		adopting \emph{only} static or symbolic execution (\emph{symex}) and
		their difference.}
	\begin{tabular}{l|rrrrrrr|rrr|rrr}
		\toprule

		\multirow{2}{*}{Use case} &
		\multirow{2}{*}{\# func.} &
		\multicolumn{2}{c}{\emph{action}} &
		\multicolumn{2}{c}{edge} &
		\multicolumn{1}{c}{\% \emph{action}} &
		\multicolumn{1}{c}{\# func.} &
		\multicolumn{3}{|c}{analysis time [s]} &
		\multicolumn{3}{|c}{trade-off \emph{actions} explored} \\

		& & \multicolumn{1}{c}{$\mu$} & \multicolumn{1}{c}{$\sigma$} &
		\multicolumn{1}{c}{$\mu$} & \multicolumn{1}{c}{$\sigma$} &
		\multicolumn{1}{c}{explored} & \multicolumn{1}{c}{static} &
		\multicolumn{1}{|c}{$\mu$} & \multicolumn{1}{c}{$\sigma$} &
		\multicolumn{1}{c}{total} & \multicolumn{1}{|c}{\emph{static}}
		& \multicolumn{1}{c}{\emph{symex}} & \multicolumn{1}{c}{$\Delta$(\%)} \\

		\midrule
		\textsf{Contact}~\cite{signalrepo} & $71$ & $12.77$ & $12.59$ & $15.09$
		& $17.64$ & $96.4\%$ & $1$ & $20.20$ & $85.9$ & $1397.12$ & $1042$ &
		$998$ & $4.41$ \\
		\textsf{libdvdcss}~\cite{libdvdcss} & $56$ & $18.50$ & $18.98$ &
		$23.84$ & $26.06$ & $91.4\%$ & $9$ & $70.19$ & $179.65$ & $3790.19$ &
		$904$ & $747$ & $21.02$ \\
		\textsf{StealthDB}~\cite{stealthdb} & $44$ & $18.29$ & $13.53$ &
		$21.97$ & $18.05$ & $96.6\%$ & $0$ & $6.16$ & $24.5$ & $258.89$ & $967$
		& $1009$ & $-4.16$ \\
		\textsf{SGX-Biniax2}~\cite{bauman2016case} & $49$ & $8.55$ &
		$8.75$ & $9.29$ & $11.71$ & $91.6\%$ & $4$ & $52.46$ & $168.8$ &
		$2465.62$ & $451$ & $413$ & $9.20$ \\
		\textsf{Unit-test} & $17$ & $6.88$ & $7.47$ & $7.17$ & $10.52$ &
		$94.0\%$ & $0$ & $15.60$ & $53.4$ & $234.29$& $122$ & $107$ & $14.02$
		\\
		\midrule
		\emph{total} & $237$ & - & - & - & - & - & $14$ & - & - & $8146.11$&
		$3486$ & $3274$ & $6.48$ \\
		\bottomrule
	\end{tabular}
	\label{tbl:coverage}
\end{table*}

\subsubsection{Model Extractor}
\label{sssec:coverage}

We analyze the Model Extractor (Section~\ref{ssec:model-exctraction}).
Specifically, we measure coverage and precision.

\textbf{Use cases complexity:}
As stated in introduction, we assume the enclave's code is \emph{simple}
enough
to be modeled with a combination of symbolic execution and static
analysis (Section~\ref{ssec:model-exctraction}).
The concept of \emph{simple enclave} has already appeared in previous
works~\cite{251582,carefulpacking}, however, they did not provide comparable
metrics.
In Table~\ref{tbl:preliminary_analysis}, we show a set of metrics that
describe the software analyzed in our use cases.
Specifically, we indicate the line of code (LoC), the number of
secure functions, and the cyclomatic complexity~\cite{ebert2016cyclomatic}.
We additionally measure the control-flow graph for each enclave's function and
report the average (and standard deviation) number of nodes and edges per
function.
Similar metrics have been previously used to indicate the effectiveness of
symbolic execution to explore a piece of software~\cite{203634}.
Finally, we count the number of direct and indirect function calls as the most
important for the security guarantee.
Intuitively, the less indirect calls an enclave has, the less likely an
adversary can carry out a mimicry attack
(\eg COOP~\cite{schuster2015counterfeit}).
One may argue that, since we assume an enclave with few indirect calls,
then bound checks can effectively stop the memory corruption attacks.
However, previous works~\cite{251582} showed that a compromised OS can input
\emph{malicious} pointers to internal enclave structures.
This allows an adversary to overwrite internal enclave data structures even
with boundary checks in place.
Therefore, using only bounds checks do not eradicate the problem in SGX
enclaves, even for \emph{simple} ones.

\textbf{Coverage:}
In the context of \sgxmonitor{}, the \emph{action} coverage is a suitable
metric for estimating the quality of an extracted model.
This comes from two observations.
First, assuming a sound symbolic execution, if no timeout is reached
(\eg $10$ minutes), we can state the analysis covered meaningful \emph{actions}.
We measure this with the percentage of traversed \emph{actions} (over 91.4\% in
our experiments).
Conversely, if the symbolic execution times out, we fallback to an
insensitive static analysis.  This traverses all the CFG of a function,
thus completing the exploration of the \emph{actions}. Of course, being the
analysis insensitive, we trade-off precision for a low
overhead in the construction of the model: we might observe rogue
\emph{actions}, which potentially increase the attack's surface.

Table~\ref{tbl:coverage} shows our coverage results. We applied the analysis
described in Section~\ref{ssec:model-exctraction} to our uses
cases: \textsf{Contact}, \textsf{libdvdcss}, \textsf{StealthDB},
\textsf{SGX-Biniax2}, and the \textsf{unit-test}.
The five use cases show a varying degree of complexity; \textsf{Contact}
contains the highest number of
single functions ($71$) among our use cases that are however quite
simple ($12$ \emph{actions} on average). Conversely,
\textsf{StealthDB} has fewer ($44$) but more
complex ($18$ \emph{actions} on average) functions.
\textsf{libdvdcss} and \textsf{SGX-Biniax2} have a
complexity similar to \textsf{StealthDB} ($18.29$ and $8.55$
\emph{actions} on average, respectively).
Finally, the \textsf{unit-test} is self-contained and primarily
leveraged to validate \sgxmonitor{} and
\emph{exception handling} of enclaves.
Overall, our analysis covers from $91.4\%$ to $96.6\%$ of the \emph{actions}.

\textbf{Precision:}
We want to inspect if the unexplored \emph{actions} caused by symbolic
execution timeout may cause false positives.
To this end, we extract three models for each use case, namely: \emph{symex},
by using only symbolic execution and interrupting the exploration once reached
timeout; \emph{static}, by using only insensitive static analysis; and
\emph{symex+static}, which is the one described in
Section~\ref{ssec:model-exctraction}.
Using only \emph{symex} models, two secure functions in \textsf{Contact}
generate \emph{false positives}, this due to the function \texttt{crecip} that
was not explored completely.
Moreover, we observe similar cases in \textsf{SGX-Biniax2} and
\textsf{libdvdcss}, in which critical functions for crypting/decrypting were
not correctly explored with only \emph{symex}.
We register false positives also using \emph{static} models, in
particular, one secure function in \textsf{StealthDB} gave false positive
because of a \texttt{jmp} not correctly resolved (see the previous paragraph).
Finally, \emph{symex+static} models did not generate any \emph{false
positive} when compared with all our tests,
thus showing that the combination of \emph{symex+static}
can significantly model the enclave behavior.
Specifically, we stress \textsf{libdvdcss}, \textsf{StealthDB}, and
\textsf{SGX-Biniax2} with long macro-benchmarks (see
Section~\ref{sssec:macro-benchmar}).
For \textsf{Contact} and the \textsf{unit-test}, we first run our
micro-benchmarks, without observing any false positives. Then, we also manually
investigated the cause of the unexplored \emph{actions}.
In most of the cases, pruned \emph{actions} are corner cases that never happen
in real executions (\eg a function that tests a null-pointer that never
happens).

Notably, the exception handler mechanism of Intel SGX SDK always introduces a
few non-traversed \emph{actions}.
This is caused by the routine \texttt{internal\_handle\_exception} that relies
on a list of pointers created at runtime.
Our Model Extractor automatically infers this structure and
resolves the indirect call in \texttt{internal\_handle\_exception} (further
details in Appendix~\ref{sec:sgx-sdk-exception-handler}).
Therefore, our Model Extractor automatically prunes those paths that never
appear at runtime, \ie if the enclave does not contain custom handlers, it will
never execute part of \texttt{internal\_handle\_exception}.

To sum up, our precision analysis shows that the
combination of the symbolic execution and the insensitive static analyses
achieve no false positives in our use cases, \ie  there are no legal
\emph{actions} that are erroneously flagged as an instance of an attack.

\noindent\begin{minipage}{\columnwidth}%
\vspace{2mm}
	\begin{mdframed}[style=HighlightFrame]\
\textbf{Model Extractor---Take Away.}
Our results show that
\begin{enumerate*}[label=(\roman*)]
	\item the symbolic execution is suitable to cover the small
	functions in SGX enclaves (\ie only $14$ functions out of $237$
	($5.9\%$) required an insensitive static analysis) and effectively
	cuts out unused \emph{actions} thus reducing the attack surface;
	\item the static analysis can support the symbolic one in case of
	timeout;
	\item our approach is practical since it can be completed in around
	an hour (\ie $60$m for \textsf{libdvdcss}); and
	\item our analysis explores a significant portion of the code since
	it does not rise false positive alarms.
\end{enumerate*}
	\end{mdframed}
\end{minipage}

%% file: relatedworks.tex
\section{Related Works}
\label{sec:related-works}

\sgxmonitor{} shares common points with different research areas.
Here, we discuss previous
provenance analysis works (Section~\ref{ssec:provenance-analysis}),
runtime RA schema (Section~\ref{ssec:runtime-remote-attestation}), and finally,
SGX and memory-corruptions (Section~\ref{ssec:sgx-memory-corr}).

\subsection{Provenance Analysis}
\label{ssec:provenance-analysis}
Many provenance tools are based on instrumentation to collect specific logs from
diverse sources~\cite{ma2016protracer,lee2013high,ma2017mpi}.
\sgxmonitor{} applies provenance to a novel area, we gather information from an isolated enclave
while the analysis runs in an zero-trust environment.
We overcome this issue with a novel technique to collect enclave runtime
fine-grain information in the presence of a malicious OS.

Other provenance techniques focus on long term intrusion, such as
APT~\cite{han_unicorn_2020,noauthor_alchemist_nodate}.
In our scenario, instead, we focus on code-reuse attacks that affect SGX enclaves.
\sgxmonitor{} helps an analyst to rebuild the intrusion by leveraging on a novel model
suited for enclaves.
\sgxmonitor{} shares some similarities with runtime provenance
works~\cite{pasquier_runtime_2018} that rely on a healthy OS to collect and analyze logs.
Conversely, \sgxmonitor{} assumes a malicious OS that might tamper with these operations.

Overall, \sgxmonitor{} is the first provenance analysis suitable for the challenging
SGX environment, providing runtime provenance analysis.
To achieve this, we design a novel log collection and propose a novel model to represent
the normal behavior of an enclave.


\subsection{Runtime Remote Attestation}
\label{ssec:runtime-remote-attestation}
Extracting and verifying runtime information remotely is similar to
Runtime Remote Attestation
works~\cite{abera2016c,aberadiat,koutroumpouchos2019secure,scarr}.
However, \sgxmonitor{} underlies different assumptions and goals compared to these works.
First, runtime remote attestation works are meant to detect intrusion at 
runtime.
Conversely, \sgxmonitor{} aims at collecting and information that could be
analyzed later on.
Since, all the previous works assume having an isolated trusted anchor to
inspect the target (\ie the enclave).
This is not possible in SGX since this feature is precluded by design.
We overcome this limitation with a novel pure software design.
Finally, the model employed by previous works are not suitable for SGX enclaves.
\sgxmonitor{}, instead, uses a novel model to capture and describe the enclave 
execution.

In GuaranTEE~\cite{morbitzer2022guarantee}, the authors propose a runtime 
attestation for SGX. However, their model is stateless and cannot
identify advanced malware such as SnakeGX.
On the contrary, both model and design of SgxMonitor are designed to cover a 
broader attacker model, moreover, we performed a more comprehensive security 
evaluation.

\subsection{SGX and Memory Corruption Errors}
\label{ssec:sgx-memory-corr}
CFIs and shadow stacks~\cite{cficc2019, hu2018enforcing, kleen2015intel,
7924286, ding2017efficient} are orthogonal defenses to \sgxmonitor{} and
complement the protection of enclaves.
In addition, one can remove corruptions errors in SGX enclaves, as studied in several
forms~\cite{kuvaiskii2017sgxbounds,schuster2015vc3,sgxrust,251582,mishra2021sgxpecial}.
All these works can be considered orthogonal to \sgxmonitor{} since they contribute
to reduce the attack surface.
However, these solutions do not provide information about the intrusion.
\sgxmonitor{}, instead, helps one rebuild the cause of an attack.

%% file: conclusion.tex
\section{Conclusion}
\label{sec:conclusion}

We proposed \sgxmonitor{}, a novel provenance analysis for SGX enclaves. As 
enclaves are designed to secure code that performs specific security- and 
privacy-sensitive tasks, \sgxmonitor{} relies on a combination of symbolic 
execution and static analysis to model the expected behavior of enclaves with 
high code coverage and low false positives.  
Moreover, \sgxmonitor{} designs a novel protocol to securely extract runtime 
enclave information in the presence of an adversarial OS.


We assessed \sgxmonitor{} security properties against novel SGX code-reuse 
attacks.
Moreover, we tested \sgxmonitor{} across four real use cases (\ie
\textsf{Contact}, \textsf{StealthDB}, \textsf{libdvdcss},
\textsf{SGX-Biniax2}) and a \textsf{unit test} to validate enclaves' corner cases.  

\sgxmonitor{} overhead is similar to the state-of-the-art provenance analysis
works showing low macro-benchmark overhead and high precision with 96\% code
coverage and zero false positives support \sgxmonitor{} in realistic deployments
to extract insight about runtime anomalous executions of SGX enclaves.

%% file: appendix.tex
\section{Model Examples}
\label{sec:model-examples}

In this section, we discuss the application of \sgxmonitor{} model 
(Section~\ref{sec:model}) over two important Intel SGX SDK mechanisms:
the outside function interaction (Section~\ref{ssec:ocall-example}) and the 
exception handling (Section~\ref{ssec:exception-handling}).

\begin{figure*}[t]
\centering
\begin{subfigure}[b]{0.43\textwidth}
	\centering
	\begin{tabular}{ll}
		\toprule 
		\textbf{Transaction} & \textbf{Definition} \\ \midrule		
		\texttt{AEX} & \emph{handled at microcode level} \\
		$T^\text{THD1}$ & $[(\text{N}, \texttt{src}, 
		\texttt{idx})_{\texttt{idx} = -3}]$ \\
		$T^\text{THD2}$ & $P \cup [(\text{J}, \texttt{src}, \texttt{ctx})]$ \\
		$T^\text{THD3}$ & $P \cup [(\text{T}, \texttt{src}, \oslash)]$ \\
		$T^\text{ERESUME}$ & $P \cup [(\text{R}, \texttt{src}, \oslash)]$ \\
		$T^\text{IHD1}$ & $P \cup [(\text{K}, \texttt{src}, \texttt{ctx})]$ \\
		$T^\text{IHD2}$ & $P \cup [(\text{J}, \texttt{src}, \texttt{ctx})]$ \\
		$T^\text{CONT}$ & $P \cup [(\text{K}, \texttt{src}, \texttt{ctx})]$ 
		\\		
		\bottomrule
	\end{tabular} 
	\caption{Transaction definition of \sgxmonitor{} model for the exception 
		handling interaction.}
	\label{tbl:transactions-exception}
\end{subfigure}
\hfill
\begin{subfigure}[b]{0.56\textwidth}
	\centering
	\includegraphics[width=\linewidth]{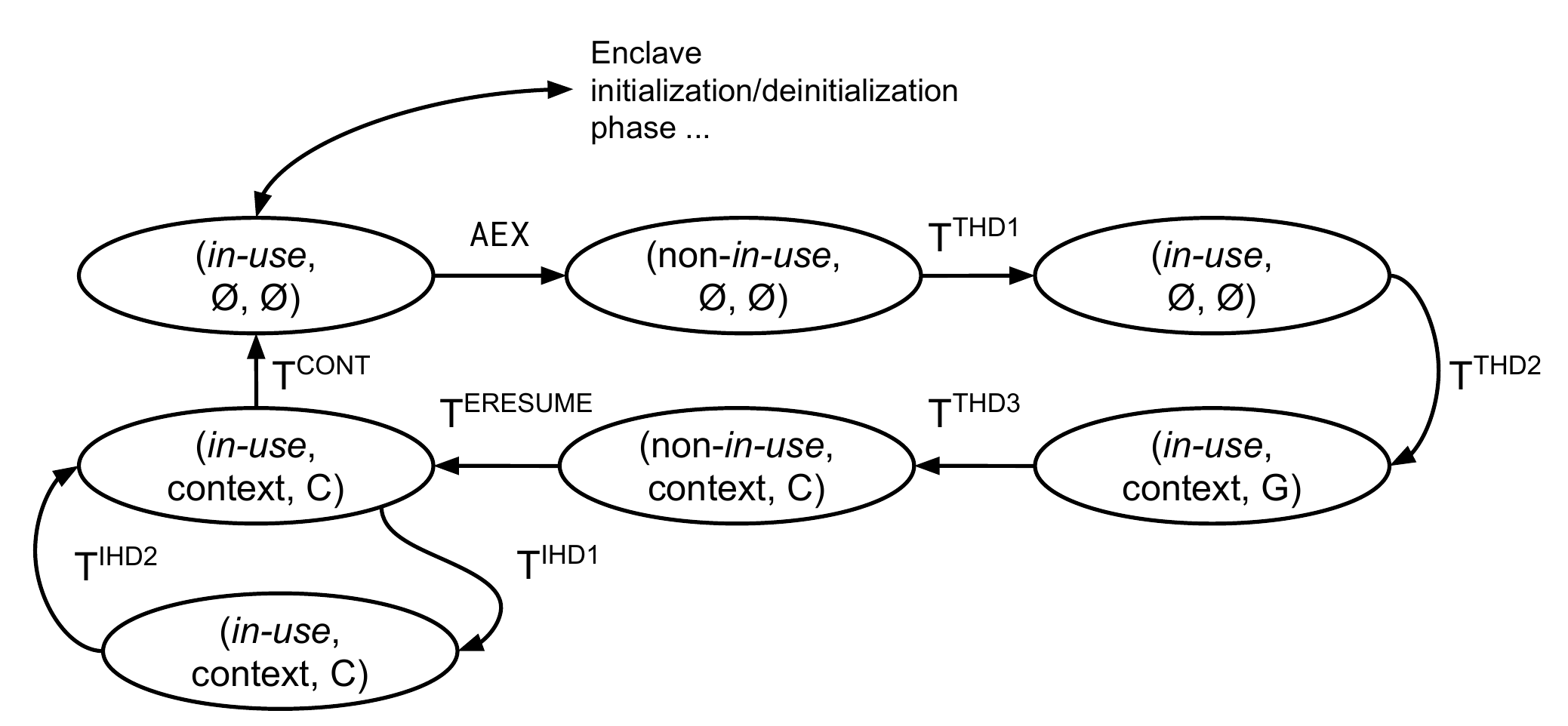}
	\caption{\sgxmonitor{} representation of exception handling.}
	\label{fig:my-model-exception}
\end{subfigure}
\caption{Example of \emph{exception handling} modeling.
	We show the FSM representation and the transaction definitions, 
	respectively.}
\label{fig:exception-handling}
\end{figure*}

\paragraph*{Transaction syntax}
For the sake of simplicity, we indicate the transactions in 
tables~\ref{tbl:transactions} and~\ref{tbl:transactions-exception} with the 
following syntax:
$$
T = P \cup [s].
$$
$T$ is composed of any \emph{valid} sequence of \emph{generic actions} $P$ 
(according to the specification of Section~\ref{sec:model}) that terminates 
with the \emph{stop action} $s$.
In case $T$ does not contain any \emph{generic action}, we omit $P$.

\subsection{Outside Function Modeling}
\label{ssec:ocall-example}

\begin{figure*}[t]
	\centering
	\begin{subfigure}[b]{0.43\textwidth}
		\centering
		\begin{tabular}{ll}
			\toprule 
			\textbf{Transaction} & \textbf{Definition} \\ \midrule
			$T^\text{ECALL}$ & $[(\text{N}, \texttt{src}, 
			\texttt{idx})_{\texttt{idx} \geq 0}]$ \\ 
			$T^\text{ERET}$ & $P \cup [(\text{T},\texttt{src},\oslash)]$ \\
			$T^\text{OCALL1}$ & $P \cup [(\text{G}, \texttt{src}, 
			\texttt{ctx})]$ \\ 
			$T^\text{OCALL2}$ & $P \cup [(\text{D},\texttt{src},\oslash)]$ 
			\\ 
			$T^\text{ORET1}$ & $[(\text{N}, \texttt{src}, 
			\texttt{idx})_{\texttt{idx} = -2}]$ \\
			$T^\text{ORET2}$ & $P \cup [(\text{C}, \texttt{src}, 
			\texttt{ctx})]$ \\
			\bottomrule
		\end{tabular} 
		\caption{Transaction definition of \sgxmonitor{} model for the 
		\emph{outside function} interaction.}
		\label{tbl:transactions}
    \end{subfigure}
	\hfill
	\begin{subfigure}[b]{0.56\textwidth}
		\centering
		\includegraphics[width=\linewidth]{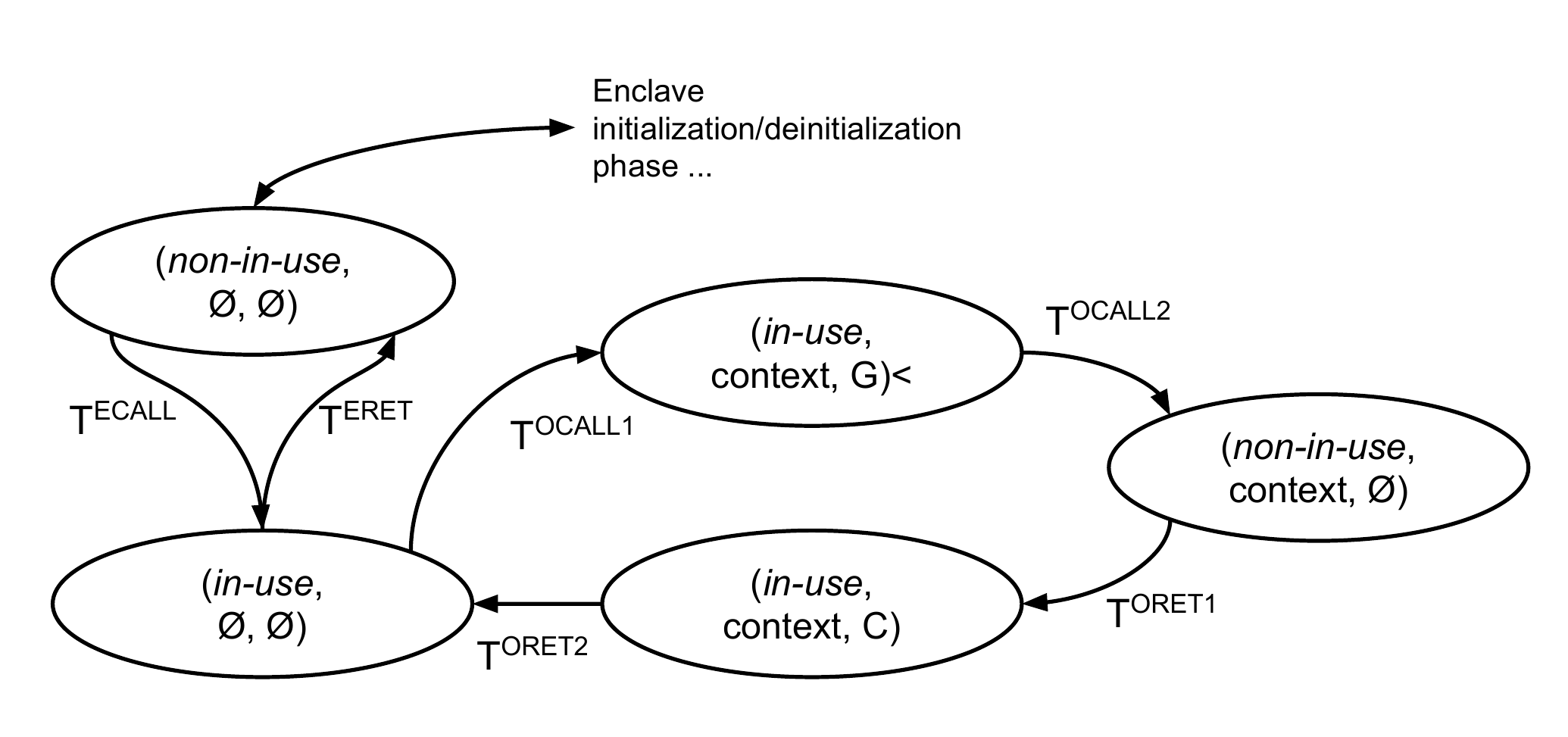}
		\caption{\sgxmonitor{} representation of \emph{outside functions} 
		interaction.}
		\label{fig:my-model}
	\end{subfigure}
	\caption{Example of \emph{outside functions} interaction modeling.
	We show the FSM representation and the transaction definitions, 
	respectively.}
	\label{fig:outside-function}
\end{figure*}

Figure~\ref{fig:outside-function} shows the application of \sgxmonitor{} to the 
enclave \emph{outside function} interaction.

After the enclave initialization, the host invokes a \emph{secure 
function}, which activates an \texttt{EENTER} opcode with the 
\texttt{idx} greater or equal than \emph{zero} (\ie $T^\text{ECALL}$).
From this point, the \emph{secure function} can evolve in two ways:
\begin{enumerate*}[label=(E\arabic*)]
	\item it does not need any interaction with the host, thus it performs an 
	ERET; or 
	\item it requires an interaction with the host, thus it performs an ORET.
\end{enumerate*}
In case (E1), the enclave does not generate any context and, therefore,
it performs a valid execution path that ends with an 
\texttt{EEXIT} opcode (\ie $T^{\text{ERET}}$).
In case (E2), instead, we need two steps to accomplish an OCALL:
\begin{enumerate*}[label=(\roman*)]
	\item generating an \texttt{ocall\_context} (\ie $T^\text{OCALL1}$), and
	\item invoking the \emph{outside function} (\ie $T^\text{OCALL2}$).
\end{enumerate*}

Once the \emph{outside function} needs to resume the \emph{secure function} 
execution, it invokes an ORET, that is composed of two steps: 
\begin{enumerate*}[label=(\roman*)]
	\item the execution enters in the enclave (\ie $T^\text{ORET1}$), and
	\item the \texttt{ocall\_context} is restored (\ie $T^\text{ORET2}$).
\end{enumerate*}
From this point ahead, the \emph{secure function} can exit the enclave
through an ERET (E1) or perform further OCALLs (E2).


\subsection{Exception Handling Modeling}
\label{ssec:exception-handling}

In Figure~\ref{fig:my-model-exception}, we depict the \sgxmonitor{} 
representation 
of the SGX SDK exception handling.
Overall, the SGX SDK handles exceptions in two phases, called \emph{trusted 
handle} (TH) and \emph{internal handle} (IH), respectively.
In the first phase (TH), the SGX interrupts its 
execution as a result of an \texttt{AEX}, and passes the control to the host.
As soon as an exception is triggered, the microcode saves the CPU
registers in a dedicated page, called SSA, for later 
stages~\cite{costan2016intel}.
After an \texttt{AEX}, the SDK expects the invocation of a dedicated 
\emph{secure function}, called \texttt{trts\_handle\_exception}, which index is 
$-3$ (\ie T$^\text{THD1}$).
This function fills an \texttt{sgx\_exception\_info\_t} structure with the 
values previously stored in the SSA (\ie T$^\text{THD2}$).
At the end of (TH), the enclave is ready for the second phase (IH) and thus it
leaves the control to the host (\ie T$^\text{THD3}$).
The host invokes an \texttt{ERESUME} to activate
the \texttt{internal\_handle\_exception} routine (\ie T$^\text{ERESUME}$).
Now, the enclave iterates among the custom handlers eventually registered
(\ie T$^\text{IHD1}$ and T$^\text{IHD2}$).
Each custom handler attempts at fixing the exception by analyzing the 
\texttt{sgx\_exception\_info\_t}, possibly altering it.
Therefore, we update the enclave internal state at each iteration.
After invoking all the internal handlers, the SGX SDK uses the
\texttt{continue\_execution} routine to resume the \emph{secure function} (\ie 
T$^\text{CONT}$).
Finally, if the exception is properly handled, the \emph{secure function} 
will continue, otherwise, a new \texttt{AEX} happens and the 
exception workflow starts again.

\subsection{SGX SDK Exception Handling}
\label{sec:sgx-sdk-exception-handler}

In the following, we show an example of 
registration of a custom exception handler, that happens by invoking the 
function \texttt{sgx\_register\_exception\_handler}.
The enclave passes the address of the exception handler as an argument, \eg 
\texttt{divide\_by\_zero\_handler}.
The Model Extractor (Section~\ref{ssec:model-exctraction}) parses the enclave 
code and identifies all the \texttt{sgx\_register\_exception\_handler} 
invocations.
Then, it performs a taint analysis to infer the address of the custom exception 
handler passed as second parameter to 
\texttt{sgx\_register\_exception\_handler}.
Finally, it uses this information to build a symbolic structure that will be 
used to explore the function \texttt{internal\_handle\_exception}, that 
actually dispatches the exception to the correct handler, if any.

\begin{lstlisting}[style=CStyle]
if (sgx_register_exception_handler(1, divide_by_zero_handler) == NULL) {
   printf("register failed\n");
} else {
   printf("register success\n");
}
\end{lstlisting}